\documentclass[lettersize,journal]{IEEEtran}
\usepackage{amsmath,amsfonts}
\usepackage{booktabs}
\usepackage{algorithmic}
\usepackage{algorithm}
\usepackage{array}
\usepackage[caption=false,font=normalsize,textfont=sf]{subfig}
\usepackage{textcomp}
\usepackage{stfloats}
\usepackage{url}
\usepackage{verbatim}
\usepackage{graphicx}
\usepackage{cite}
\usepackage{multicol}
\usepackage{amssymb}
\usepackage{lineno}
\usepackage{makecell}
\usepackage{indentfirst}
\usepackage{bm}
\usepackage{color}
\usepackage{cite}
\usepackage{epstopdf}
\usepackage{amssymb}
\usepackage{xcolor}
\usepackage{tikz}
\usepackage{mathtools}
\usepackage{pifont}
\newcommand{\diag}{\mathrm{diag}}

\newcommand{\tr}{\mathrm{tr}}
\newcommand{\HH}{\mathrm{H}}	
\newcommand{\TT}{\mathrm{T}}
\newtheorem{theorem}{Theorem}

\newtheorem{remark}{Remark}

\begin{document}
	
	\title{Model-Driven Sensing-Node Selection and Power Allocation for Tracking Maneuvering Targets in Perceptive Mobile Networks}
	
	\author{Lei Xie,~\IEEEmembership{Member,~IEEE}, Hengtao He,~\IEEEmembership{Member,~IEEE}, Shenghui Song,~\IEEEmembership{Senior Member,~IEEE},\\ and Yonina C. Eldar,~\IEEEmembership{Fellow,~IEEE}
		\thanks{L. Xie, H. He, and S. Song are with Department of Electronic and Computer Engineering, the Hong Kong University of Science and Technology, Hong Kong. e-mail: ($\{$eelxie, eehthe, eeshsong$\}$@ust.hk). Y. C. Eldar is with the Faculty of Mathematics and Computer Science, Weizmann Institute of Science, Rehovot 7610001, Israel (e-mail: yonina.eldar@weizmann.ac.il).}
	}
	
	
	
	\maketitle
	
	\begin{abstract}
    Maneuvering target tracking will be an important service of future wireless networks to assist innovative applications such as intelligent transportation. However, tracking maneuvering targets by cellular networks faces many challenges. For example, the dense network and high-speed targets make the selection of the sensing nodes (SNs) and the associated power allocation very challenging. Existing methods demonstrated engaging performance, but with high computational complexity. In this paper, we propose a model-driven deep learning (DL)-based approach for SN selection. To this end, we first propose an iterative SN selection method by jointly exploiting the majorization-minimization (MM) framework and the alternating direction method of multipliers (ADMM). Then, we unfold the iterative algorithm as a deep neural network and prove its convergence. The proposed method achieves lower computational complexity, because the number of layers is less than the number of iterations required by the original algorithm, and each layer only involves simple matrix-vector additions/multiplications. Finally, we propose an efficient power allocation method based on fixed point (FP) water filling and solve the joint SN selection and power allocation problem under the alternative optimization framework. Simulation results show that the proposed method achieves better performance than the conventional optimization-based methods with much lower computational complexity.
	\end{abstract}
	
	
	\section{Introduction}

	Innovative applications such as intelligent transportation systems require high-precision sensing capabilities, which cannot be provided by current cellular networks. To this end, the recently proposed integrated sensing and communication (ISAC) paradigm offers a promising way to share spectrum, hardware, and software between sensing and communication \cite{8288677,liu2020radar}.  Perceptive mobile network (PMN) was proposed as a special type of ISAC system that adds high-precision sensing capability to cellular networks \cite{9296833,xie2022perceptive,xie2022networked,xie2023collaborative}. There are many favorable properties of cellular networks that can facilitate sensing. For instance, the large number of sensing nodes (SNs) in PMNs enables collaborative sensing, where multiple perspectives from different SNs are exploited to sense the same target. 
	The SNs can be base station (BS) \cite{8288677}, road side units \cite{liu2020radar}, remote radio unit \cite{9296833}, or target monitoring terminal \cite{xie2022perceptive}. 

	However, tracking maneuvering targets by PMNs faces many challenges. For example, due to the dense cellular network, selecting a proper set of SNs to track a moving target can be very difficult, because the handover from one group of SNs to another faces very stringent latency requirements. There have been engaging results on SN selection and power allocation for tracking maneuvering targets \cite{macsazade2010energy,7104065,yan2020optimal,shen2013sensor,yi2020resource,yan2015simultaneous,yuan2020robust,xie2017joint,sun2021resource}.
	In \cite{macsazade2010energy}, two SN selection methods in wireless networks were proposed to minimize the posterior Cram\'{e}r-Rao lower bound (PCRLB) and maximize the mutual information between the target location and the measurements of the selected SNs, respectively. In \cite{7104065}, a cooperative game theoretic approach was utilized to allocate power for tracking targets in a radar network.
	In \cite{yan2020optimal}, two strategies for resource allocation with given SNs were proposed, where one maximizes the tracking accuracy with limited power budgets, and the other minimizes the power consumption with required tracking performance. 
	
	To achieve better performance, the joint SN selection and power allocation schemes were also considered \cite{xie2017joint,sun2021resource}. 
	In \cite{xie2017joint}, a distributed multi-target tracking method was proposed for the networked multiple-input multiple-output (MIMO) radar system, where an alternative optimization (AO)-based method was utilized to solve the bi-variable optimization problem. The boolean constraint on the SN selection vector is one of the most critical challenges for the joint SN selection and power allocation problem.
	To handle this issue, a typical method is to relax the boolean constraint to allow continuous and sparse variables \cite{das2011submodular,elhamifar2015dissimilarity,sun2021resource}.	
	In \cite{xie2017joint,sun2021resource}, the relaxed SN selection was formulated as a semi-definite programming (SDP) problem and solved by the CVX toolbox \cite{grant2014cvx}. 
	Unfortunately, the complexity of the existing methods increases exponentially with the number of SNs, which may violate the stringent latency requirement of sensing applications when a large number of SNs exist.

	To this end, model-driven deep learning (DL) offers a promising solution \cite{he2019model,he2020model}. 
	By unfolding an iterative algorithm as a neural network where each iteration is implemented by one layer with learnable parameters, model-driven methods have the potential to offer better performance with reduced computational complexity.
	Some research efforts have been made to utilize model-driven deep neural networks (DNNs) to find sparse solutions for better performance and lower computational costs. 
	In \cite{borgerding2017amp}, an unfolded vector-approximate message passing network with random initialization was proposed to learn a denoiser identical to the statistically matched one. In \cite{xin2016maximal}, the iterative algorithm, designed to solve a problem with $l_0$ sparse regularization, was unfolded to be a feed-forward neural network for faster inference and better scalability. In \cite{8550778}, a generalized DNN was proposed to learn a sparse solution by unfolding the alternating direction method of multipliers (ADMM) with better accuracy and lower computational cost. In \cite{9420308}, an ADMM-Net is designed for interference removal in radar imaging, which exhibited much lower imaging error and computational cost than ADMM and CVX. 
	However, the inverse of high-dimensional matrices are involved in the existing ADMM-based unfolding methods, which causes high storage and computational cost. 
	
	In this paper, to meet the stringent latency requirement of sensing applications, we propose a model-driven method for SN selection to track multiple maneuvering targets. For that purpose, we first derive an iterative algorithm for SN selection, leveraging the majorization-minimization (MM) framework and ADMM. 
	Then, the MM-ADMM algorithm is unfolded into a DNN where the technical challenges lie in the large number of learnable parameters and the uncertain convergence property. To this end, we design a new model-driven DNN with an additional module to exploit the first- and second-order momentum, and refer to it as deep alternating network (DAN), which has fewer learnable parameters than the directly-unfolded MM-ADMM.
	The convergence proof of the proposed DAN is also given. The computational complexity of DAN is low, because the number of layers is less than the number of iterations required by the original algorithm, and each layer of DAN only involves simple matrix-vector additions/multiplications without high-dimensional matrix inverse. 
	Finally, we propose a fixed-point (FP) water-filling (WF)-based method for power allocation, which is derived based on the Lagrange multiplier method. 
	The joint SN selection and power allocation problem is solved by combining the proposed DAN and FP-WF algorithms under the AO framework. Experiment results show that the proposed method can achieve better performance than the optimization-based methods with remarkably lower computational costs.
	
	The contributions of this paper are summarized as follows:
	\begin{enumerate}
		\item We propose an iterative method based on MM and ADMM for SN selection. In particular, we exploit the MM approach to handle the non-convexity of the penalized cost functions. For each iteration of ADMM, we derive explicit expressions for the solution to the constrained optimization problem by exploiting the Karush–Kuhn–Tucker (KKT) conditions, which facilitate the development of the model-driven method. 
		\item We design a new model-driven DNN, named DAN, by adding an additional module to the directly-unfolded MM-ADMM method, which exploits the momentum for accelerating the convergence. 
		Moreover, we provide the convergence proof for DAN, which achieves a similar SN selection performance as the exhaustive searching method with significantly lower computational cost.
		\item Inspired by the classic WF-based power allocation strategies, we propose an iterative FP-WF power allocation method. Specifically, in each water-filling step, the water level is obtained by solving an FP equation. 
		This approach not only reduces the computational complexity, but also provides an interesting physical insight: the power allocation strategy depends on the ratio between the Fisher information of the predictions and the measurements.
	\end{enumerate}
	
	The remainder of this paper is organized as follows. Section II introduces the system model and formulates the problem. Section III derives the joint SN selection and power allocation algorithm. Section IV provides the simulation results to validate the advantage of the proposed model-driven method. Section V concludes this paper.
	
	\section{System Model and Problem Formulation}
\begin{figure}[!t]
			\centering
			\includegraphics[width=3.0in]{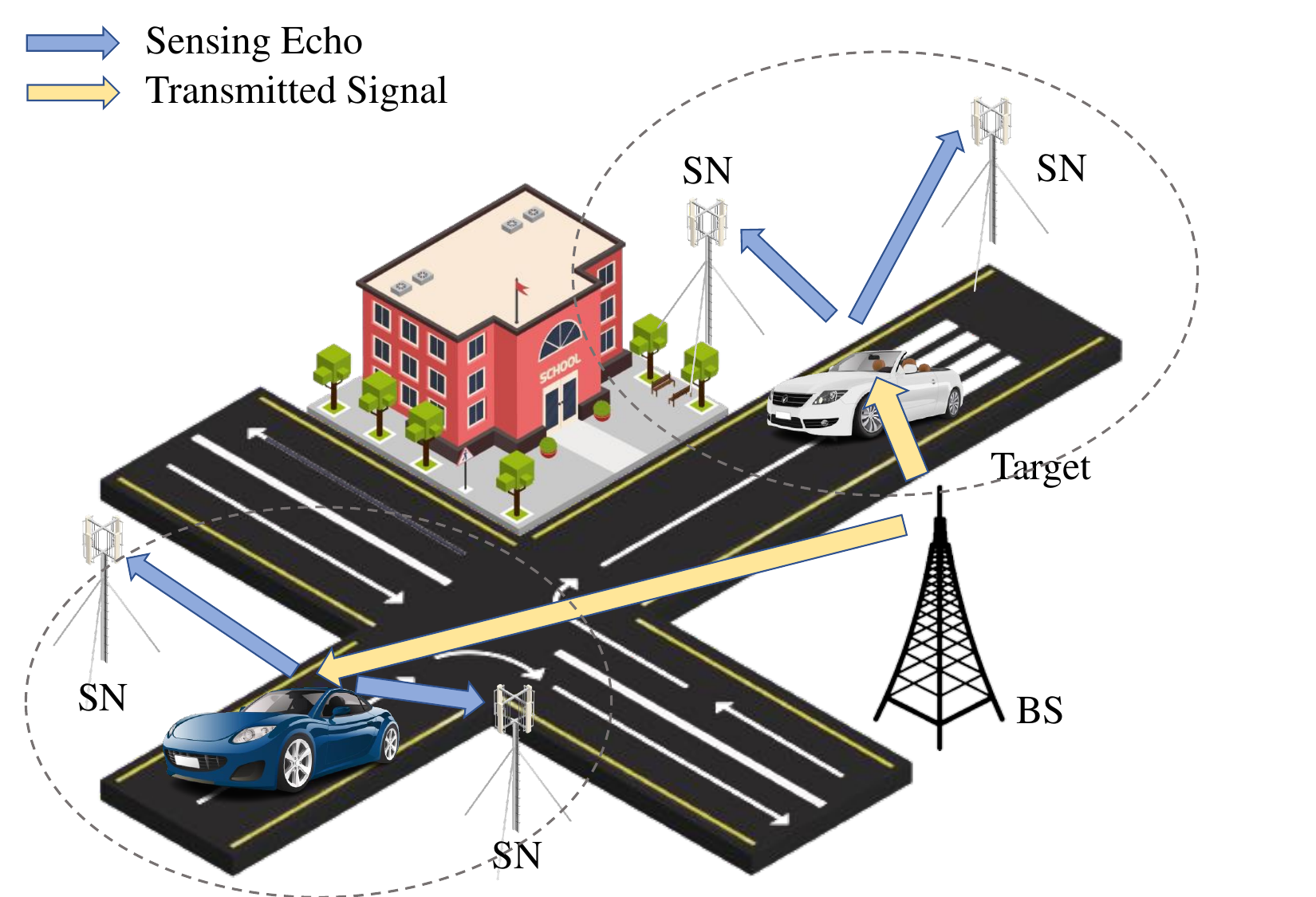}
			\caption{Illustration of the system. }
			\label{fig_sys}
		\end{figure}

	In Fig. \ref{fig_sys}, we show a PMN consisting of one BS serving as the sensing signal transmitter and $N$ SNs serving as the receivers for the echoes, which can be BSs or other types of SNs \cite{8288677,liu2020radar,9296833,xie2022perceptive}. In each tracking frame, the BS will transmit sensing signals to the predicted positions of multiple targets, and the selected SNs will collaboratively estimate the location and velocity of the targets (motion state). The estimation results will be utilized to predict the motion state in the next tracking frame. 
	In this paper, the SN selection and power allocation will be formulated as an optimization problem to minimize the PCRLB for the estimation error of the target motion state. To this end, we first introduce the target motion model and the signal model, which are necessary for deriving the PCRLB. 


		
		\subsection{Target Motion Model}
		The target motion model describes the motion behavior of the targets and affects the Fisher information of the prediction.
		Assume that the target motion follows a near constant velocity model and the transition matrix $\mathbf{G}$ is given by \cite{yan2015simultaneous,yi2020resource,yuan2020robust,sun2021resource}
		\begin{equation}
			\begin{split}
				\mathbf{G}=\mathbf{I}_2\otimes \left[\begin{matrix}
					1 & \Delta T\\
					0 & 1\\
				\end{matrix}\right]
			\end{split}
		\end{equation}
		where $\mathbf{I}_2$ denotes the $2\times 2$ identity matrix, $\otimes$ represents the Kronecker product, and $\Delta T$ denotes the time between two adjacent tracking frames.
		In the $k$th tracking frame, there are $Q$ point-like targets, where the $q$th target is located at  $\mathbf{r}_{q}^{(k)}=(r_{x,q}^{(k)},r_{y,q}^{(k)})$ with a velocity $\mathbf{v}_{q}^{(k)}=(v_{x,q}^{(k)},v_{y,q}^{(k)})$. 
		The target motion state is updated by $\mathbf{x}_{q}^{(k)} = \mathbf{G} \mathbf{x}_{q}^{(k-1)}+ \mathbf{z}_{q}^{(k-1)}$,
		where 
		$\mathbf{x}_q^{(k)} = [r_{x,q}^{(k)},v_{x,q}^{(k)},r_{y,q}^{(k)},v_{y,q}^{(k)}]^\TT$ includes the parameters to be estimated.
		Here, $\mathbf{z}_q^{(k-1)}$ denotes the state noise, which is assumed to be a zero-mean Gaussian vector with covariance matrix  \cite{yan2015simultaneous,yi2020resource}
		\begin{equation}
			\begin{split}
				\mathbf{Q}=q_s \mathbf{I}_2\otimes \left[\begin{matrix}
					\frac{1}{3}(\Delta T)^3 & \frac{1}{2}(\Delta T)^2\\
					\frac{1}{2}(\Delta T)^2 & \Delta T\\
				\end{matrix}\right]
			\end{split}
		\end{equation}
		where $q_s$ is the intensity of the process noise. 	We assume the number of the targets is known in the previous frame, and the targets are widely separated and each of them moves independently in the monitoring area \cite{yuan2020robust}.  
		%
		%
		%
		%
		%
		%
		%
		\subsection{Signal Model}
		In the $k$th tracking frame, the BS will transmit the sensing signal  $\mathbf{s}^{(k)}(t)$ to the targets, and the echoes will be captured by the selected SNs for sensing purposes. The location of the BS and the $n$th SN is given by $\mathbf{r}_{BS}$ and $\mathbf{r}_n$, respectively.  
		Given the motion state, we can determine the measurements, i.e., the angle of arrival (AOA), the time delay, and the Doppler frequency of the $q$-th target with respect to the $n$-th SN as
		\begin{align}
			&\theta_{q,n}^{(k)} =\arccos\frac{\mathbf{e}_{n}^\TT(\mathbf{r}_{q}^{(k)}-\mathbf{r}_{n})}{\Vert\mathbf{r}_{q}^{(k)}-\mathbf{r}_{n}\Vert},\\
			&\tau_{q,n}^{(k)}=\frac{1}{c}\left(\Vert\mathbf{r}_n-\mathbf{r}_{q}^{(k)}\Vert+\Vert\mathbf{r}_{BS}-\mathbf{r}_{q}^{(k)}\Vert\right),\\
			&\mu_{q,n}^{(k)}=\frac{\mathbf{v}_{q}^\TT(\mathbf{r}_{q}^{(k)}-\mathbf{r}_{n})}{\lambda\Vert\mathbf{r}_{q}^{(k)}-\mathbf{r}_{n}\Vert}    +  \frac{\mathbf{v}_{q}^\TT(\mathbf{r}_{q}^{(k)}-\mathbf{r}_{BS})}{\lambda\Vert\mathbf{r}_{q}^{(k)}-\mathbf{r}_{BS}\Vert} ,
		\end{align}
		where $\mathbf{e}_n$ represents the unit vector parallel to the line formed by all antennas of the uniform linear array, $c$ is the speed of light, $\lambda$ is the wavelength, and $||\cdot||$ denotes the $l_2$ norm.

		Define the power allocation vector $\mathbf{p}^{(k)}=[p_1^{(k)},\cdots,p_Q^{(k)}]\in \mathbb{R}^{Q\times 1}$, where $p_q^{(k)}$ denotes the power allocated to the $q$th target. 
		The baseband echo of the $q$th target received by the $n$th SN is given by
		\begin{equation}\label{yqnt}
			\begin{split}
				\mathbf{y}_{q,n}^{(k)}(t)&=\sqrt{p_q^{(k)}}\beta_{q,n}^{(k)} e^{j2\pi \mu_{q,n}^{(k)}t} \mathbf{b}_{q,n}^{(k)}\mathbf{a}_{q,k}^\HH \mathbf{s}^{(k)}(t-\tau_{q,n}^{(k)})\\
				&\quad +\mathbf{n}_{n}^{(k)}(t),
			\end{split}
		\end{equation}
		where $\mathbf{n}_{n}^{(k)}(t)$ denotes the complex additive white Gaussian noise with zero mean and variance $\sigma^2$. The transmit and receive steering vectors are given by  $\mathbf{b}_{q,n}^{(k)}=\mathbf{b}(\theta_{q,n}^{(k)})$ and $\mathbf{a}_{q,k}=\mathbf{a}(\psi_q^{(k)})$, respectively, where $\psi_q^{(k)}$ represents the angle of departure (AOD) of the $q$th target from the BS. 	$\beta_{q,n}^{(k)}$ represents the complex gain of the BS-target-SN ($q$th target and $n$th SN) path, which accounts for the array gain, the propagation loss and the radar cross section (RCS) \cite{xie2020recursive}. 
			Following \cite{yan2015simultaneous,yi2020resource,yuan2020robust,xie2017joint,sun2021resource}, the local estimation error is modeled as a zero-mean Gaussian vector with the covariance matrix
			\begin{equation}\label{meacov}
				\begin{split}
					\mathbf{\Sigma}_{q,n}^{(k)}=\diag\left[\sigma_{\theta_{q,n}^{(k)}}^2,\sigma_{\tau_{q,n}^{(k)}}^2,\sigma_{\mu_{q,n}^{(k)}}^2\right],
				\end{split}
			\end{equation}
			where  $\sigma_{\theta_{q,n}^{(k)}}^2$, $\sigma_{\tau_{q,n}^{(k)}}^2$, and $\sigma_{\mu_{q,n}^{(k)}}^2$ denote the CRLBs for the estimation of the direction, range, and Doppler shift, respectively. The local estimation error affects the Fisher information of measurement, which will be utilized to derive the PCRLB in the next section.

			\subsection{Posterior Cram\'{e}r-Rao Lower Bound}
			Based on the above-mentioned target motion model and signal model, we will derive the PCRLB, which gives the lower bound of the estimation error for the target motion state. 
			Define $\mathbf{U}^{(k)}=[\mathbf{u}_1^{(k)},\cdots,\mathbf{u}_Q^{(k)}]\in \mathbb{R}^{N_{BS}\times Q}$ as the SN selection matrix, whose $(n,q)$th entry $u_{q,n}^{(k)}$ is 1 if the $q$th target is associated with the $n$th SN. 
			The Fisher information matrix (FIM) for the $q$th target is given by \cite{7181639}
			\begin{equation}\label{FIMdef}
				\mathbf{J}_q^{(k)}(p_q^{(k)},\mathbf{u}_q^{(k)})=\mathbf{J}_{P,q}^{(k)}+\mathbf{J}_{Z,q}^{(k)},
			\end{equation}
			where $\mathbf{J}_{P,q}^{(k)}$ and $\mathbf{J}_{Z,q}^{(k)}$ denote the prior and data information matrix, respectively. In particular, the prior information matrix is given by
			\begin{equation}
				\mathbf{J}_{P,q}^{(k)}=\left(\mathbf{Q}+\mathbf{G}	(\mathbf{J}_q^{(k-1)})^{-1}\mathbf{G}^{\HH}\right)^{-1}.
			\end{equation}
			The data information matrix $\mathbf{J}_{Z,q}^{(k)}$ is given by
			\begin{equation}
				\mathbf{J}_{Z,q}^{(k)}=\sum_{n=1}^{N} u_{q,n}^{(k)}(\mathbf{H}_{q,n}^{(k)})^\TT (\mathbf{\Sigma}_{q,n}^{(k)})^{-1} \mathbf{H}_{q,n}^{(k)},
			\end{equation}
			where 
			\begin{equation}
				\mathbf{H}_{q,n}^{(k)}=
				\frac{\partial \mathbf{g}_{n}^{(k)}}{\partial \mathbf{x}_q^{(k)}}
				\bigg\vert_{\mathbf{x}_q^{(k)}=\hat{\mathbf{x}}_q^{(k|k-1)}},
			\end{equation}
			with $\frac{\partial \mathbf{g}_{n}^{(k)}}{\partial \mathbf{x}_q^{(k)}}$ denoting the derivative of the measurements $\mathbf{g}_{n}^{(k)}=[\theta_{q,n}^{(k)}(\mathbf{x}_{q}^{(k)}),\tau_{q,n}^{(k)}(\mathbf{x}_{q}^{(k)}),\mu_{q,n}^{(k)}(\mathbf{x}_{q}^{(k)})]^\TT$ with respect to the motion state $\mathbf{x}_q^{(k)}$. 
		The predicted motion state of the $q$th target in the $k$th frame is updated by $\hat{\mathbf{x}}_q^{(k|k-1)}=\mathbf{G}\hat{\mathbf{x}}_q^{(k-1)},$ 
		where $\hat{\mathbf{x}}_q^{(k-1)}$ represents the estimated motion state of the $q$th target in the $(k-1)$th frame. 
		Note that $\mathbf{\Sigma}_{q,n}^{(k)}$ is inversely proportional to the SNR at the SN \cite{yan2015simultaneous,yi2020resource,yuan2020robust,xie2017joint,sun2021resource}. 
		Thus, we can rewrite the measurement covariance in (\ref{meacov}) as 
		\begin{equation}
			\mathbf{\Sigma}_{q,n}^{(k)}
			=(p_{q}^{(k)})^{-1}\bar{\mathbf{\Sigma}}_{q,n}^{(k)},
		\end{equation}
		where $\bar{\mathbf{\Sigma}}_{q,n}^{(k)}$ contains the part of $\mathbf{\Sigma}_{q,n}^{(k)}$ that is independent of $p_{q}^{(k)}$.
		Then, we have $\mathbf{J}_{Z,q}^{(k)}=
		p_q^{(k)}
		\sum_{n=1}^{N} u_{q,n}^{(k)} \overline{\mathbf{M}}_{q,n}^{(k)}$, 
		where 
		$\overline{\mathbf{M}}_{q,n}^{(k)}=(\mathbf{H}_{q,n}^{(k)})^\TT (\bar{\mathbf{\Sigma}}_{q,n}^{(k)})^{-1} \mathbf{H}_{q,n}^{(k)}$. 
		Note that $\mathbf{M}_{q,n}^{(k)}=p_q^{(k)}\overline{\mathbf{M}}_{q,n}^{(k)}$ denotes the measurement information for the $q$th target at the $n$th SN. 
		The inverse of the derived FIM yields the PCRLB matrix, i.e., \cite{yan2015simultaneous,yi2020resource,yuan2020robust,xie2017joint,sun2021resource}
		\begin{equation}
			\mathbf{C}_q(p_q^{(k)},\mathbf{u}_q^{(k)})=\left(\mathbf{J}_q^{(k)}(p_q^{(k)},\mathbf{u}_q^{(k)})\right)^{-1}.
		\end{equation}
		The diagonal elements of $\mathbf{C}_q(p_q^{(k)},\mathbf{u}_q^{(k)})$ provide a lower bound on the variances of the estimation error of an unbiased estimator for the target motion state, i.e.,
		\begin{equation}
			\mathbb{E}\left((\hat{\mathbf{x}}_q^{(k)}-\mathbf{x}_q^{(k)})(\hat{\mathbf{x}}_q^{(k)}-\mathbf{x}_q^{(k)})^{\HH}\right)\succeq \mathbf{C}_q(p_q^{(k)},\mathbf{u}_q^{(k)}),
		\end{equation}
		where 
		$\mathbf{A}\succeq\mathbf{B}$ indicates  $\mathbf{A}-\mathbf{B}$ is a positive-semidefinite matrix. 
		Some functions of the diagonal elements of the PCRLB matrix, e.g., the trace \cite{yan2015simultaneous} and the determinant \cite{325008}, have been used as performance metrics for target sensing and tracking. 
		
		%
		
		\subsection{Problem Formulation}
		In this paper, we will minimize PCRLB through SN selection and power allocation. In the $k$th frame, the optimization problem is modeled as 
		\begin{subequations}\label{Problem0}	
			\begin{align}
				\min_{\mathbf{p}^{(k)},\mathbf{U}^{(k)}} \;&\sum_{q=1}^Q \log \det \mathbf{C}_q(p_q^{(k)},\mathbf{u}_q^{(k)}) \notag\\
				\text{s.t.} \quad &\sum_{q=1}^Q p_q^{(k)} \leq P_T, \label{cont00}\\
				& p_q^{(k)} \geq P_{\min},\label{cont001}\\
				&\mathbf{1}^\TT \mathbf{u}_q^{(k)}\leq N_{\max},q=1,2,\cdots,Q,  \label{cont01}\\
				&\mathbf{U}^{(k)}\in\{0,1\}^{N\times Q}, \label{cont02}
			\end{align}	
		\end{subequations}
		where constraint (\ref{cont00}) limits the total transmit power. Constraint  (\ref{cont001}) indicates the minimum power allocated to each target,  constraint (\ref{cont01}) limits the maximum number of SNs to track one target \cite{bejar2001distributed}, and  (\ref{cont02}) gives the binary constraint on $\mathbf{u}_q^{(k)}$. The main reasons to select $\log \det(\mathbf{C}_q)$ as the performance metric include: 1) the determinant of $\mathbf{C}_q$ is proportional to the volume of the minimum achievable covariance ellipsoid, which is widely used as an important metric for parameter estimation \cite{325008}; and 2) if the determinant is directly used, the original problem (\ref{Problem0}) is not convex, but the monotonic logarithmic transformations can render this problem convex.

		\section{Model-Driven DL-based Sensing Node Selection and Power Allocation Scheme}
		Note that the problem in (\ref{Problem0}) has two optimization variables. To handle this issue, we propose to update the variables alternatively based on the AO theory.
		With a given feasible starting point  $\left\{ {\mathbf{p}}^{(k,0)}, \{{\mathbf{u}}_q^{(k,0)}\}_{q=1}^Q \right\}$, we iteratively perform the following two operations: 
		
		1) updating $\{{\mathbf{u}}_q^{(k,j+1)}\}_{q=1}^{Q}$ with fixed ${\mathbf{p}}^{(k,j)}$ via
		\begin{equation}\label{subj2}
			\begin{split}
				\mathbf{u}_q^{(k,j+1)}=&\arg \min_{\mathbf{u}_q^{(k)}}  \log \det \mathbf{C}_q(p_q^{(k,j)},\mathbf{u}_q^{(k)}),
			\end{split}
		\end{equation}
		
		2) updating ${\mathbf{p}}^{(k,j+1)}$ with fixed $\{{\mathbf{u}}_q^{(k,j+1)}\}_{q=1}^{Q}$ via
		\begin{equation}\label{subj1}
			\mathbf{p}^{(k,j+1)}=\arg \min_{\mathbf{p}^{(k)}} \sum_{q=1}^Q \log \det \mathbf{C}_q(p_q^{(k)},\mathbf{u}_q^{(k,j+1)}),
		\end{equation}
		which decouple the joint SN selection and power allocation problem. 
		
		In the following, we will first develop an iterative method for SN selection by jointly exploiting the MM framework and ADMM.
		To further reduce the computational complexity, we will utilize the model-driven DL technique to solve (\ref{subj2}). 	
		Finally, we will propose an FP-based WF method to solve (\ref{subj1}), which has much lower complexity but offers comparable performance as the traditional CVX-based method.

		\subsection{MM-ADMM based Sensing Node Selection}
		Given $\mathbf{p}^{(k,j)}$, the problem in  (\ref{subj2}) can be formulated as
		\begin{equation}\label{prob0u}
			\begin{split}
				\min_{\mathbf{u}_q^{(k)}} \;&\mathcal{F}_u(\mathbf{u}_q^{(k)})\\
				\text{s.t.} \; &\mathbf{1}^\TT \mathbf{u}_q^{(k)}\leq N_{\max},\;\mathbf{u}_q^{(k)}\in\{0,1\}^{N\times 1},
			\end{split}
		\end{equation}
		where $\mathcal{F}_u(\mathbf{u}_q^{(k)})=\log \det \mathbf{C}_q(\mathbf{u}_q^{(k)}|p_{q}^{(k,j)})$. 
		In order to enforce a binary solution and simplify the problem, we introduce a $l_0$ pseudo-norm penalty to the objective function and relax the binary constraint \cite{zhai2018joint}. Then, the problem in (\ref{prob0u}) is relaxed as
		\begin{equation}\label{prob01u}
			\begin{split}
				\min_{\mathbf{u}_q^{(k)}} \;&\mathcal{F}_u(\mathbf{u}_q^{(k)})+\rho_q\Vert\mathbf{u}_q^{(k)}\Vert_0\\
				\text{s.t.} \; &\mathbf{1}^\TT \mathbf{u}_q^{(k)}\leq N_{\max},\;\mathbf{0}\leq \mathbf{u}_q^{(k)}\leq \mathbf{1},
			\end{split}
		\end{equation}
		where $\Vert\cdot\Vert_0$ denotes the $l_0$ pseudo-norm.
		In general, a larger $\rho_q$ leads to a sparser $\mathbf{u}_q^{(k)}$. Due to the non-convex, non-continuous, and combinatorial nature of the $l_0$ pseudo-norm, the problem (\ref{prob01u}) is NP-hard. To simplify the notation, we omit the index $q$ hereafter unless doing so creates confusion.

		Inspired by \cite{malek2016successive}, we approximate the $l_0$ pseudo-norm by a function $\mathcal{P}_\gamma(\mathbf{u}^{(k)})=\sum_{n=1}^{N}(1-e^{-\gamma u_n^{(k)}})$, 
		where $\gamma$ is a sufficiently large constant. $\mathcal{P}_\gamma(\mathbf{u}^{(k)})$ is utilized due to several favorable properties: 1) it is asymptotically equivalent to $\Vert\mathbf{u}^{(k)}\Vert_0$, i.e.,
		$\lim_{\gamma\to \infty}\mathcal{P}_\gamma(\mathbf{u}^{(k)})=\sum_{n=1}^{N}(1-\delta(u_n^{(k)}))=\Vert\mathbf{u}^{(k)}\Vert_0$;
		2) it is continuous, concave, and non-decreasing in the feasible set; and	3) it is differentiable and its gradient is easy to obtain.

		\subsubsection{MM framework for solving (\ref{prob01u})}
		The problem in (\ref{prob01u}) can be approximated by
		\begin{equation}\label{prob1u}
			\begin{split}
				\min_{\mathbf{u}^{(k)}\in \mathcal{S}_u} \;&\mathcal{F}_u(\mathbf{u}^{(k)})+\rho \mathcal{P}_\gamma(\mathbf{u}^{(k)})
			\end{split}
		\end{equation}
		where  $\mathcal{S}_u=\{\mathbf{u}^{(k)}|\mathbf{1}^\TT \mathbf{u}^{(k)}= N_{\max},\mathbf{0}\leq \mathbf{u}^{(k)}\leq \mathbf{1}\}$. 
		Though  $\mathcal{P}_\gamma(\mathbf{u}^{(k)})$ is continuous w.r.t. $\mathbf{u}^{(k)}$, the problem in (\ref{prob1u}) is still hard to solve, due to the complicated form of $\mathcal{F}_u(\mathbf{u}^{(k)})$ w.r.t. $\mathbf{u}^{(k)}$. To handle this difficulty, we propose to utilize the MM framework \cite{sun2016majorization}, based on which (\ref{prob1u}) can be solved in an iterative process. At each iteration, the  MM framework updates the optimization variable by minimizing a tight upperbound of the function, which is known as the surrogate function. 
		Then, the next question is how to construct a surrogate function for the objective function in (\ref{prob1u}). 
		
		Since $\mathcal{P}_\gamma(\mathbf{u}^{(k)})$ is differentiable and concave with respect to $\mathbf{u}^{(k)}$, it is upperbounded by its first-order Taylor expansion, i.e.,
		\begin{equation}\label{prob2u}
			\begin{split}
				&\mathcal{P}_\gamma(\mathbf{u}^{(k)})\leq \widetilde{\mathcal{P}}_\gamma(\mathbf{u}^{(k)}|\mathbf{u}^{(k,l)})\\
				&\triangleq\mathcal{P}_\gamma(\mathbf{u}^{(k,l)}) + (\mathbf{d}_{\gamma}^{(k,l)})^\TT (\mathbf{u}^{(k)}-\mathbf{u}^{(k,l)}),\\
			\end{split}
		\end{equation}
		where $\mathbf{u}^{(k,l)}$ denotes the optimized result at the $l$th iteration,  $\mathbf{d}_{\gamma}^{(k,l)}=\gamma[e^{-\gamma u_1^{(k,l)}},e^{-\gamma u_2^{(k,l)}},\cdots,e^{-\gamma u_{N}^{(k,l)}}]^\TT$ represents the gradient of $\mathcal{P}_\gamma(\mathbf{u}^{(k)})$, and $u_n^{(k,l)}$
		denotes the $n$th entry of $\mathbf{u}^{(k,l)}$.

		%
		%
		
		An appropriate upperbound of $\mathcal{F}_u(\mathbf{u}^{(k)})$ can be obtained by
		\begin{align}
			& \widetilde{\mathcal{G}}_1(\mathbf{u}^{(k)}|\mathbf{u}^{(k,l)})\triangleq\mathcal{F}_u(\mathbf{u}^{(k,l)})+\mathbf{d}_{u}^\TT(\mathbf{u}^{(k,l)})(\mathbf{u}^{(k)}-\mathbf{u}^{(k,l)})\notag\\
			&\quad+\frac{1}{2}(\mathbf{u}^{(k)}-\mathbf{u}^{(k,l)})^\TT\mathbf{T}^{(k,l)}(\mathbf{u}^{(k)}-\mathbf{u}^{(k,l)}),
		\end{align}
		where $\mathbf{d}_u^{(k,l)}=	\mathbf{d}_u(\mathbf{u}^{(k,l)})$ and $	\mathbf{d}_u(\mathbf{u}^{(k)})=\frac{\partial \mathcal{F}_u(\mathbf{u}^{(k)})}{\partial \mathbf{u}^{(k)}}$
		denotes the gradient of $\mathcal{F}_u(\mathbf{u}^{(k)})$ w.r.t. $\mathbf{u}^{(k)}$, whose $n$th entry is given by $d_{u,n}(\mathbf{u}^{(k)})=\frac{\partial \mathcal{F}_u(\mathbf{u}^{(k)})}{\partial u_n^{(k)}}=-\tr \left((\mathbf{J}^{(k)}(\mathbf{u}^{(k)}|p^{(k,j)}))^{-1}\mathbf{M}_{n}^{(k)}\right).$
		The positive-definite matrix $\mathbf{T}^{(k,l)}$ should satisfy 
		\begin{equation}
			\mathbf{T}^{(k,l)}\succeq \mathbf{H}_{u}(\mathbf{u}^{(k,l)}),
			\label{Tcondition}
		\end{equation}
		where $\mathbf{H}_u(\mathbf{u}^{(k)})=\frac{\partial \mathcal{F}_u(\mathbf{u}^{(k)})}{\partial \mathbf{u}^{(k)} \partial(\mathbf{u}^{(k)})^\TT}$ 
		denotes the Hessian matrix of $\mathcal{F}_u(\mathbf{u}^{(k)})$ w.r.t. $\mathbf{u}^{(k)}$, whose $(m,n)$th entry is given by $H_{u,m,n}(\mathbf{u}^{(k)})=\frac{\partial \mathcal{F}_u(\mathbf{u}^{(k)})}{\partial u_m^{(k)}\partial u_n^{(k)}}=\tr\; \mathbf{M}_{m}^{(k)}\left(\mathbf{J}^{(k)}(\mathbf{u}^{(k)}|p^{(k,j)})\right)^{-2}\mathbf{M}_{n}^{(k)}$. 
		Then, at the $(l+1)$th iteration, the selection vector can be updated by solving the problem
		\begin{equation}\label{prob1Gu}
			\begin{split}
				\min_{\mathbf{u}^{(k)}\in \mathcal{S}_u} \;&\mathcal{G}(\mathbf{u}^{(k)}),\\
			\end{split}
		\end{equation}
		where the surrogate function $\mathcal{G}(\mathbf{u}^{(k)})$ is defined by
		\begin{equation}\label{surro}
			\begin{split}
				\mathcal{G}(\mathbf{u}^{(k)})=\widetilde{\mathcal{G}}_1(\mathbf{u}^{(k)}|\mathbf{u}^{(k,l)})+\rho\widetilde{\mathcal{P}}_\gamma(\mathbf{u}^{(k)}|\mathbf{u}^{(k,l)}).
			\end{split}
		\end{equation}
		The problem in (\ref{prob1Gu}) is convex and can be solved by the general CVX toolbox based on the interior point method (IPM) \cite{grant2014cvx}. However, the computational complexity of CVX is about $\mathcal{O}(N^{3.5})$, which is not suitable for PMNs with a large $N$.
		
		\subsubsection{ADMM-based method for solving (\ref{prob1Gu})}
		To solve (\ref{prob1Gu}) efficiently, we exploit the ADMM, which splits the problem into two distinct parts and handles them separately \cite{boyd2011distributed}. Since (\ref{surro}) is Lipschitz continuous, the convergence of the ADMM can be guaranteed.
		By introducing an auxiliary variable $\mathbf{v}^{(k)}$, (\ref{prob01u}) is equivalent to
		\begin{equation}\label{prob02u}
			\begin{split}
				\min_{\mathbf{u}^{(k)},\mathbf{v}^{(k)}} \;&\widetilde{\mathcal{G}}_1(\mathbf{u}^{(k)}|\mathbf{u}^{(k,l)})+\rho\widetilde{\mathcal{P}}_\gamma(\mathbf{v}^{(k)}|\mathbf{u}^{(k,l)})\\
				\text{s.t.} \quad &\mathbf{1}^\TT \mathbf{u}^{(k)}= N_{\max},\;\mathbf{0}\leq \mathbf{v}^{(k)}\leq \mathbf{1},\;\mathbf{u}^{(k)}=\mathbf{v}^{(k)},
			\end{split}
		\end{equation}
		which leads to the augmented Lagrangian function \cite{boyd2011distributed}
		\begin{align}
			\mathcal{L}(\mathbf{u}^{(k)},\mathbf{v}^{(k)},\mathbf{z}^{(k)})&=\widetilde{\mathcal{G}}_1(\mathbf{u}^{(k)}|\mathbf{u}^{(k,l)})+\rho\widetilde{\mathcal{P}}_\gamma(\mathbf{v}^{(k)}|\mathbf{u}^{(k,l)})\notag\\
			&+\frac{\rho_{a,l}}{2}\Vert\mathbf{u}^{(k)}-\mathbf{v}^{(k)}+\mathbf{z}^{(k)}\Vert^2,\label{Func_lag0}
		\end{align}
		where $\mathbf{z}^{(k)}$ is the dual variable and $\rho_{a,l}$ is a penalty parameter at the $l$th iteration.
		Then, at the $m$th iteration, the optimization variables are updated as
		\begin{subequations}\label{ADMMiteration}
			\begin{align}	
				&\mathbf{u}_{m+1}^{(k,l)} =\arg \min_{\mathbf{u}^{(k)}} \mathcal{L}(\mathbf{u}^{(k)},\mathbf{v}_{m}^{(k,l)},\mathbf{z}_{m}^{(k,l)}),\label{subprobu}\\
				&\quad \quad \text{s.t.}\; \mathbf{1}^\TT \mathbf{u}^{(k)}= N_{\max},\notag\\
				&\mathbf{v}_{m+1}^{(k,l)}=\arg \min_{\mathbf{v}^{(k)}} \mathcal{L}(\mathbf{u}_{m+1}^{(k,l)},\mathbf{v}^{(k)},\mathbf{z}_{m}^{(k,l)}),\label{subprobv}\\
				&\quad \quad \text{s.t.}\;\mathbf{0}\leq \mathbf{v}^{(k)}\leq \mathbf{1},\notag\\
				&\mathbf{z}_{m+1}^{(k,l)}=\mathbf{z}_{m}^{(k,l)}+\mathbf{u}_{m+1}^{(k+1,l)}-\mathbf{v}_{m+1}^{(k+1,l)},\label{zup}
			\end{align}
		\end{subequations}
		where $\mathbf{u}_{m}^{(k,l)}$, $\mathbf{v}_{m}^{(k,l)}$ and  $\mathbf{z}_{m}^{(k,l)}$ denote $\mathbf{u}$, $\mathbf{v}$ and $\mathbf{z}$ at the $m$th ADMM iteration, respectively.
		
		\emph{a) Update $\mathbf{u}_{m+1}^{(k,l)}$  via \eqref{subprobu}:}
		By utilizing the Lagrange multiplier method, (\ref{subprobu}) can be reformulated as an unconstrained problem, whose Lagrange function
		is given by $\mathcal{L}_{u}(\mathbf{u}^{(k)})=\mathcal{L}(\mathbf{u}^{(k)},\mathbf{v}_{m}^{(k,l)},\mathbf{z}_{m}^{(k,l)})+\nu_l(N_{\max}-\mathbf{1}^\TT \mathbf{u}^{(k)})$,
		with  a Lagrange multiplier $\nu_{l}$. The closed-form solution to (\ref{subprobu}) is
		\begin{equation}
			\begin{split}
				\mathbf{u}_{m+1}^{(k,l)}=\mathbf{u}^{(k,l)}-\mathbf{\Phi}_u^{-1}(\mathbf{d}_{m}^{(k,l)}-\nu_{l}\mathbf{1}),
			\end{split}\label{um1}
		\end{equation}
		where
		$\mathbf{\Phi}_{l}=\mathbf{T}^{(k,l)}+\rho_{a,l} \mathbf{I}$ and $	\mathbf{d}_{m}^{(k,l)}=\mathbf{d}_u^{(k,l)}-\rho_{a,l} (\mathbf{v}_{m}^{(k,l)}-\mathbf{z}_{m}^{(k,l)})$.
		By substituting (\ref{um1}) into the constraint of (\ref{subprobu}), we have
		\begin{equation}\label{nuldef}
			\nu_{l}=\frac{N_{\max}-\mathbf{1}^\TT\mathbf{u}^{(k,l)}+\mathbf{1}^\TT\mathbf{\Phi}_{l}^{-1}\mathbf{d}_{m}^{(k,l)}}{\mathbf{1}^\TT\mathbf{\Phi}_{l}^{-1}\mathbf{1}}=\frac{\mathbf{1}^\TT\mathbf{\Phi}_{l}^{-1}\mathbf{d}_{m}^{(k,l)}}{\mathbf{1}^\TT\mathbf{\Phi}_{l}^{-1}\mathbf{1}},
		\end{equation}
		which follows from the fact that $N_{\max}=\mathbf{1}^\TT\mathbf{u}^{(k,l)}$.
		Therefore, the closed-form solution to (\ref{subprobu}) is given by
		\begin{equation}
			\begin{split}
				&\mathbf{u}_{m+1}^{(k,l)}=\mathbf{u}^{(k,l)}-\mathbf{\Phi}_{l}^{-1}\left(\mathbf{d}_{m}^{(k,l)}-\frac{\mathbf{1}^\TT\mathbf{\Phi}_{l}^{-1}\mathbf{d}_{m}^{(k,l)}}{\mathbf{1}^\TT\mathbf{\Phi}_{l}^{-1}\mathbf{1}}\mathbf{1}\right).
			\end{split}\label{um2}
		\end{equation}
		One remaining problem is how to determine $\mathbf{\Phi}_{l}$, which is equivalent to choosing a proper $\mathbf{T}^{(k,l)}$.
		Indeed, it is not difficult to find a matrix $\mathbf{T}^{(k,l)}$ that satisfies (\ref{Tcondition}), such as $\mathbf{T}^{(k,l)}= \mathbf{H}_{u}(\mathbf{u}^{(k,l)})+\epsilon \mathbf{I}$, where $\epsilon$ is a  positive constant to make $\mathbf{T}^{(k,l)}$ positive definite. 
		However, the matrix inversion of $\mathbf{\Phi}_l$ is involved in (\ref{um2}) when updating $\mathbf{u}_{m+1}^{(k,l)}$, which may be computationally complex due to the large number of SNs.
		To tackle this issue, $\mathbf{T}^{(k,l)}$ is desired to be a diagonal matrix. 
		One feasible solution is to make $\mathbf{T}^{(k,l)}$ proportional to the identity matrix, i.e., \cite{XIE2020107401}
		\begin{equation}\label{Tchoice}
			\mathbf{T}^{(k,l)}=C_T^{(k,l)}\mathbf{I},
		\end{equation}
		where $C_T^{(k,l)}$ is a positive constant to satisfy (\ref{Tcondition}). For example, one feasible choice is $C_T^{(k,l)}=\lambda_{\max}\left(\mathbf{H}_{F}(\mathbf{u}^{(k,l)})\right)$ and $\lambda_{\max}(\mathbf{X})$ denotes the principle eigenvalue of $\mathbf{X}$.

		\emph{b) Update $\mathbf{v}_{m+1}^{(k,l)}$ via \eqref{subprobv}:}
		Since (\ref{subprobv}) is convex, the closed-form solution $\mathbf{v}_{m+1}^{(k,l)}$ to (\ref{subprobv}) can be obtained based on the KKT conditions, whose $n$th entry is given by
		\begin{equation}\label{vm2}
			v_{m+1,n}^{(k)}=\left\{
			\begin{matrix*}[l]
				\widetilde{v}_n,& \text{if}\; 0\leq \widetilde{v}_n\leq 1,\\
				0, & \text{if}\;  \widetilde{v}_n< 0,\\
				1, & \text{if}\;  \widetilde{v}_n> 1,\\
			\end{matrix*}
			\right.
		\end{equation}
		where $\widetilde{v}_n$ denotes the $n$th entry of $\widetilde{\mathbf{v}}$, given by
		\begin{equation}\label{uupdateMM}
			\widetilde{\mathbf{v}}=-\frac{\rho}{\rho_{a,l}}\mathbf{d}_{\gamma}^{(k,l)}+\mathbf{u}_{m+1}^{(k)}+\mathbf{z}_{m}^{(k)}.
		\end{equation}


		\begin{remark}
			The cost function will not increase over the ADMM iteration process given in  (\ref{ADMMiteration}). According to the monotone bounded theorem \cite{bibby_1974}, the iteration will converge to a set of stationary points in the feasible set, denoted by $\mathbf{u}_{(\star)}^{(k)}$, $\mathbf{v}_{(\star)}^{(k)}$, and $\mathbf{z}_{(\star)}^{(k)}$. The selection vector $\mathbf{u}^{(k,l+1)}$ is updated by $\mathbf{u}_{(\star)}^{(k)}$. 
		\end{remark}

		\begin{remark}\label{rem3}
			The convergence and performance of (\ref{um2}) depend on the selection of $\mathbf{T}^{(k,l)}$. If $\mathbf{T}^{(k,l)}$ is selected as the Hessian matrix which is usually not diagonal, (\ref{um2}) is similar to the Newton's descent update with  quadratic convergence, but high computational complexity. In (\ref{Tchoice}), $\mathbf{T}^{(k,l)}$ is selected as a diagonal matrix, i.e., $\mathbf{T}^{(k,l)}=C_T^{(k,l)}\mathbf{I}$, and thus the update in (\ref{um2}) moves in the opposite direction of the gradient, which resembles the gradient descent method. With a diagonal $\mathbf{T}^{(k,l)}$, the computational cost at each ADMM iteration is about $\mathcal{O}(N^2)$, which is much lower than that of CVX.
			In general, a larger $C_T^{(k,l)}$ is desired to satisfy (\ref{Tcondition}).
			However, in this case, 
			the constant $C_T^{(k,l)}+\rho_{a,l}$  
			is inversely proportional to the step size. An aggressive choice of $C_T^{(k,l)}$ may require more iterations to converge. Meanwhile, the choice of $\mathbf{T}^{(k,l)}$ suggested in (\ref{Tchoice}) may not be optimal, and a better one within a larger feasible set, i.e., diagonal but not necessarily proportional to the identity matrix, is desired. 
			To this end, we propose to unfold the iterative optimization method as a DNN and tune $\mathbf{T}^{(k,l)}$ with DL techniques.

			One feasible way is to treat the diagonal elements of $\mathbf{T}^{(k,l)}$ as the learnable parameters. In this case, the number of learnable parameters is $ N$ at each layer, which will be large due to the dense SNs. Moreover, the trained $\mathbf{T}^{(k,l)}$ may break the convergence condition (\ref{Tcondition}). These issues motivate us to consider another design with three desirable properties: 1) the number of learnable parameters is moderate, 2) the convergence property is guaranteed, and 3) the proposed method will be restricted to first-order methods that only require gradients, since higher-order optimization methods may cost a large amount of computing and storage resource.
		\end{remark}

		\subsection{Deep-Alternative-Network: DL-Based Sensing Node Selection}
		To derive a DNN with the above-mentioned properties, we unfold the MM-ADMM-based SN selection method and introduce an additional module. The new DNN is called DAN.  
		As shown in Fig. \ref{fig_DAN_illu}, DAN	consists of $L$ cascaded layers with some learnable parameters, where the $(l+1)$th layer takes the first- and second-order momentum $\hat{\mathbf{m}}^{(l-1)}$ and $\hat{\mathbf{v}}^{(l-1)}$, the gradients $\mathbf{d}_{u}^{(k,l)}$ and $\mathbf{d}_{v}^{(k,l)}$, and the output from the previous layer $\mathbf{u}^{(k,l)}$ as inputs, and outputs an update $\mathbf{u}^{(k,l+1)}$. In particular, the $(l+1)$th layer updates $\mathbf{u}_{m}^{(k,l)}$, $\mathbf{v}_{m}^{(k,l)}$, and $\mathbf{z}_{m}^{(k,l)}$, alternatively, as shown by the blue, green, and orange blocks in Fig. \ref{fig_DAN_illu}, respectively. The update of $\mathbf{u}_{m+1}^{(k,l)}$ is of the same form as (\ref{um2}). But we make the following two modifications, as shown by the red block in Fig. \ref{fig_DAN_illu}:
		
		\begin{figure}[t]
			\centering
			\includegraphics[width=3.01in]{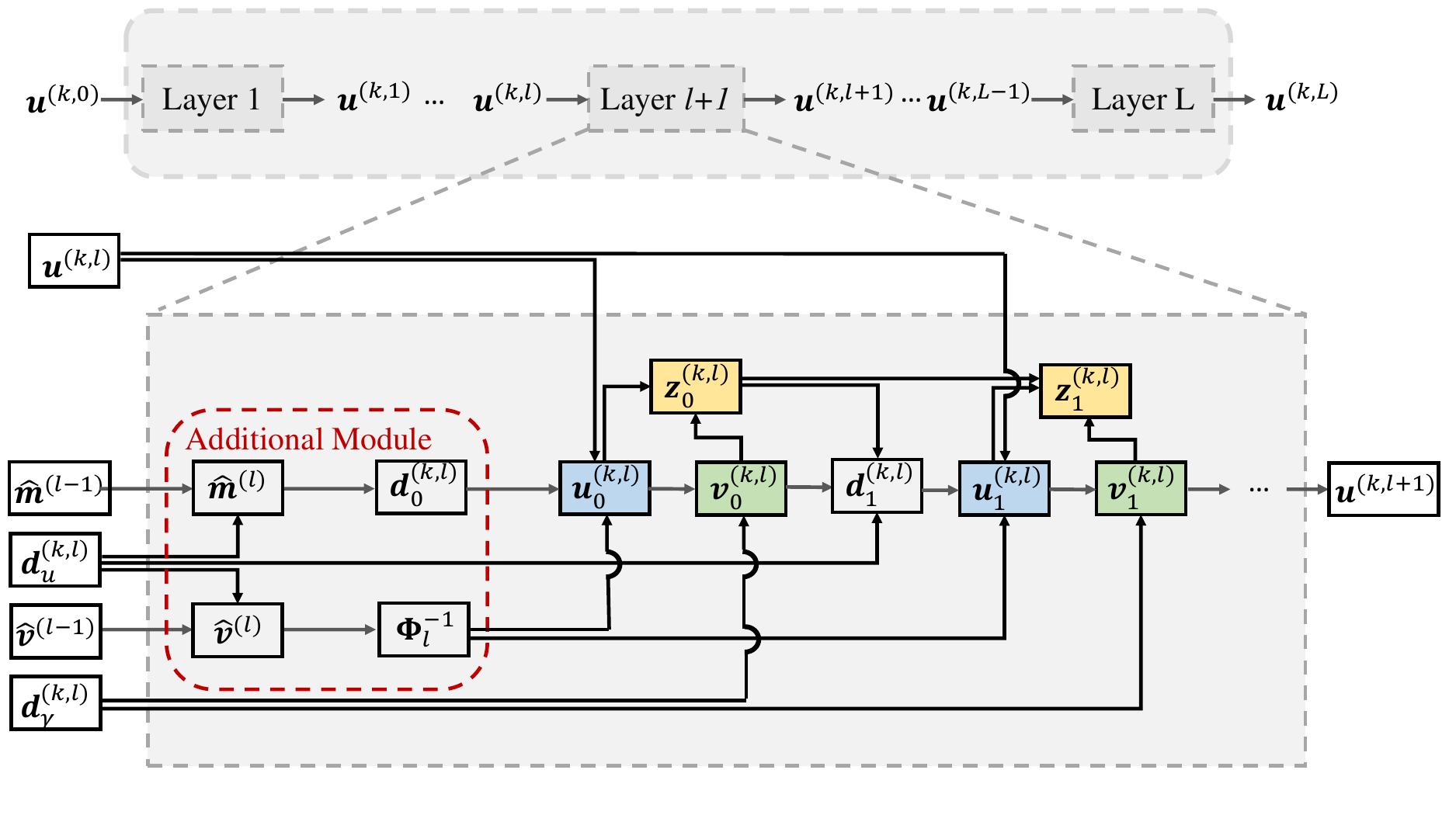}\
			\caption{Illustration of DAN.}
			\label{fig_DAN_illu}
		\end{figure}
		
		1) $\mathbf{d}_{m}^{(k,l)}$ is constructed as
		\begin{equation}\label{fsmoment}
			\begin{split}
				\mathbf{d}_{m}^{(k,l)}=\hat{\mathbf{m}}_{l}-\rho_{a,l} (\mathbf{v}_{m}^{(k,l)}-\mathbf{z}_{m}^{(k,l)}),
			\end{split}
		\end{equation}
		where
		\begin{equation}\label{firstordermom}
			\begin{split}
				\hat{\mathbf{m}}_{l}=\beta_{1,l}\hat{\mathbf{m}}_{l-1}+(1-\beta_{1,l})\mathbf{d}_u^{(k,l)}.
			\end{split}
		\end{equation}
		Here, $\beta_{1,l}=\beta_{1}\eta_{1}^l$ where $\eta_{1}\in (0,1)$ and $\beta_{1}\in (0,1)$ denotes a learnable hyper-parameters to avoid the case that the momentum diverges severely.

		When $\beta_{1,l} = 0$, the first-order momentum $\hat{\mathbf{m}}_{l}$ reduces to the gradient $\mathbf{d}_u^{(k,l)}$. In this paper, we define $\beta_{1,l}=\beta_{1}\eta_{1}^l$ with $\beta_{1}\in (0,1)$ and $\eta_{1}\in (0,1)$.
		The momentum terms caused by non-zero $\beta_{1,l}$ may improve the performance significantly, especially in DL applications.

		2) $\mathbf{\Phi}_{l}$ is constructed as 
		\begin{equation}\label{dnntm}
			\mathbf{\Phi}_{l}=\hat{\mathbf{T}}^{(k,l)}+\rho_{a,l} \mathbf{I},
		\end{equation}
		where  $\hat{\mathbf{T}}^{(k,l)}\triangleq\diag \left(\left[\frac{\sqrt{|\hat{v}_{l,1}|}}{\alpha_{1,l}},\cdots,\frac{\sqrt{|\hat{v}_{l,N}|}}{\alpha_{1,l}}\right]\right),$ and $\rho_{a,l}=\rho_{a}\eta_{a}^l$ with $\eta_{a}^l \in (0,1)$.
		Here, $\hat{v}_{l,i}$ denotes the $i$th entry of the second-order momentum $\hat{\mathbf{v}}_{l}$, which is defined by 
		\begin{equation}\label{hatvl}
			\hat{\mathbf{v}}_{l}=\beta_{2}\hat{\mathbf{v}}_{l-1}+(1-\beta_{2})(\mathbf{d}_{u}^{(k,l)})^2,
		\end{equation}
		where $\beta_{2}$ denotes a constant to control the second-order momentum
		and $\alpha_{1,l}=\frac{\bar{\alpha}_{1,l}}{\sqrt{l}}$ with $\bar{\alpha}_{1,l}\in [\alpha_{1}^{-},\alpha_{1}^{+}]$ representing a set of learnable parameters to control the update step size. Here, the positive constants $\alpha_{1}^{-}$ and $\alpha_{1}^{+}$ are the lower and upper bounds of  $\bar{\alpha}_{1,l}$. 
		We refer to the diagonal element of  $\mathbf{\Phi}_{l}^{-1}$ as the learning rate of this algorithm, whose $i$th entry is given by $\phi_{l,i}^{-1} = \left({\sqrt{|\hat{v}_{l,i}|}}/{\alpha_{1,l}} +\rho_{a,l}\right)^{-1}$. 
		Learning rate decay is critical for training neural networks. In the early training stage, a large learning rate can accelerate training and help the network escape spurious local minima. By the end of the iteration, a small learning rate helps the network converge to a local minimum and avoid oscillation. Therefore, we desire a set of $\rho_{a,l}$ and $\alpha_{1,l}$ such that, for any $ l\in \{2,\cdots,L\}$ and $i \in \{1,\cdots,N\}$, we have $\phi_{l,i}^{-1} \leq \phi_{l-1,i}^{-1}$.

		The updates are inspired by the adaptive momentum (Adam) method \cite{kingma2014adam}, i.e., an algorithm for first-order gradient-based optimization. 
		Adam is chosen due to several favorable properties: 1) simple implementation, computationally efficient, and low memory requirements; 2) adaptability to large-scale problems; and 3) adaptation to sparse gradients \cite{kingma2014adam}.
		Based on the adaptive estimates of first and second-order momentum, we propose a novel construction of $\mathbf{d}_{m}^{(k,l)}$ and $\hat{\mathbf{T}}^{(k,l)}$ as well as its resultant $\mathbf{\Phi}_{l}$, which can meet the constraint in (\ref{cont01}) and the diagonal requirement, simultaneously. 
		But different from ADAM, the update has additional terms resulting from the original MM-ADMM and one learnable step size $\alpha_{1,l}$ to control the iteration process. 
		Compared with training all diagonal elements of $\hat{\mathbf{T}}^{(k,l)}$, the learnable parameters in the DAN are changed to $\bar{\alpha}_{1,l}$ and $\beta_{1}$. The total number of learnable parameters over all layers is reduced from $L N$ to $L+1$.
		
		The update of $\mathbf{v}_{m+1}^{(k,l)}$ and $\mathbf{z}_{m+1}^{(k,l)}$ are the same as (\ref{vm2}) and (\ref{zup}), respectively. 	
		With given $\hat{\mathbf{m}}_{l}$ and $\mathbf{\Phi}_{l}$, the Lagrange function $\mathcal{L}(\mathbf{u}^{(k)},\mathbf{v}^{(k)},\mathbf{z}^{(k)}|\hat{\mathbf{m}}_{l},\mathbf{\Phi}_{l})$ defined in (\ref{Func_lag0}) will not increase after updating $\mathbf{u}_{m}^{(k,l)}$, $\mathbf{v}_{m}^{(k,l)}$ and $\mathbf{z}_{m}^{(k,l)}$ by (\ref{um2}), (\ref{vm2}), and (\ref{zup}), respectively. The modified ADMM iteration will also converge at a set of station points denoted by $\mathbf{u}_{(\star)}^{(k)}$, $\mathbf{v}_{(\star)}^{(k)}$, and $\mathbf{z}_{(\star)}^{(k)}$.
		Therefore, we have
		\begin{equation}
			\begin{split}
				&\mathbf{u}^{(k,l+1)}=\mathbf{u}_{\star}^{(k,l)}=\mathbf{u}^{(k,l)}-\mathbf{\Phi}_{l}^{-1}\left(\mathbf{d}_{\star}^{(k,l)}-\nu_{l}\mathbf{1}\right),
			\end{split}\label{um2netstar}
		\end{equation}
		where
		\begin{equation}\label{dstar}
			\mathbf{d}_{\star}^{(k,l)}=\hat{\mathbf{m}}_{l}-\rho_{a,l} (\mathbf{v}_{\star}^{(k,l)}-\mathbf{z}_{\star}^{(k,l)}),\;	\nu_{l}=\frac{\mathbf{1}^\TT\mathbf{\Phi}_{l}^{-1}\mathbf{d}_{\star}^{(k,l)}}{\mathbf{1}^\TT\mathbf{\Phi}_{l}^{-1}\mathbf{1}}.
		\end{equation}

		\subsection{Convergence of DAN}
		Until now, we have developed a new model-driven DL-based method for SN selection. 
		However, the obtained $\hat{\mathbf{T}}^{(k,l)}$ may not satisfy (\ref{Tcondition}), which indicates that the convergence property of the MM framework is questionable. 
		To address this issue, we next analyze the convergence of the proposed DAN.
		
		For any sequence $\{\mathbf{u}^{(k,l)}\}_{l=1}^L$ generated by the proposed DAN, the regret function is defined as 
		\begin{equation}
			\begin{split}
				R_L\triangleq \sum\limits_{l=1}^{L}\left( \mathcal{G}(\mathbf{u}^{(k,l)})-\mathcal{G}(\mathbf{u}^{(k,\star)})\right),
			\end{split}
		\end{equation}
		where $\mathbf{u}^{(k,\star)} =\arg \min_{\mathbf{u}^{(k)}\in\mathcal{S}_u} \mathcal{G}(\mathbf{u}^{(k)})$ denotes the best stationary point in the feasible set $\mathcal{S}_u$. Generally speaking, the regret function indicates the sum of the difference between $\mathcal{G}(\mathbf{u}^{(k,l)})$ and $\mathcal{G}(\mathbf{u}^{(k,\star)})$, which is widely used for the convergence proof \cite{kingma2014adam}. Note that the feasible set has bounded diameter, i.e., for all $\mathbf{u},\mathbf{v}\in \mathcal{S}_u$, $||\mathbf{u} - \mathbf{v}||^2 \leq D_{\Delta}$.
		Define $D_{u,1}\triangleq \max\limits_{l} ||\mathbf{d}_{u}^{(k,l)}||_1$, $D_{\phi}\triangleq \max\limits_l  \max\limits_i \phi_{l,i}^{-1}$, $D_{b,1}\triangleq\max_{l} ||\hat{\mathbf{b}}_{l}||_1$, and $D_{b,2}\triangleq\max_{l} ||\hat{\mathbf{b}}_{l}||^2$,
		where 
		$\hat{\mathbf{b}}_{l} =  \mathbf{v}_{\star}^{(k,l)}-\mathbf{z}_{\star}^{(k,l)}$.
		Then, we have the following theorem for the convergence analysis.

		\begin{theorem}
			\label{MMconvergence}
			Assume that, for all $l\in[2,L]$, $\phi_{l,i}^{-1} \leq \phi_{l-1,i}^{-1}$. 
			The regret is bounded by 
			\begin{equation}\label{regretbound}
				\begin{split}
					&R_L\leq C_1 \sqrt{L} + C_2,
				\end{split}
			\end{equation}
			where $C_1 = \frac{\sqrt{1-\beta_{2}}D_{u,1} D_{\Delta}}{\alpha_{1}^{-}(1-\sqrt{\beta_2})(1-\beta_{1})}$ and $C_2$ is defined by (\ref{regbound}), given at the top of this page.

			
		\end{theorem}
		
		\emph{Proof:} See Appendix \ref{MMconvergenceProof}.
		
		Theorem \ref{MMconvergence} indicates that the DAN has a regret of $\mathcal{O}(L^{\frac{1}{2}})$, which guarantees that the sequence $\{\mathcal{G}(\mathbf{u}^{(k,l)})\}_{l=1}^L$ will converge to $\mathcal{G}(\mathbf{u}^{(k,\star)})$ with convergence rate on the order of $\mathcal{O}(L^{-\frac{1}{2}})$.

		\begin{figure*}[!t]
			\normalsize
			\begin{align}
				&C_2=  \frac{2\rho_{a}  D_{\Delta}\eta_{a}}{1-\beta_{1}}+ \frac{\sqrt{1-\beta_{2}}D_{u,1} D_{\Delta}}{\alpha_{1}^{-}(1-\sqrt{\beta_2})(1-\beta_{1})}    + \frac{\alpha_{1}^{+}(3+\beta_{1})D_{u,1}}{2(1-\beta_{1})^2(1-\frac{\beta_{1}}{\sqrt{\beta_{2}}})\sqrt{1-\beta_{2}}(1-\eta_{1}^2)}+\frac{\rho_a  D_{\phi} (D_{b,1}+D_{b,2})}{2(1-\eta_a)(1-\beta_{1})}\notag \\
				&+ \frac{ D_{u,1}D_{\phi}}{2(1-\eta_1)(1-\beta_1)^2} + \frac{\beta_{1}\sqrt{1-\beta_{2}}D_{u,1} D_{\Delta}}{2\alpha_{1}^{-}(1-\sqrt{\beta_2})(1-\eta_{1})^2(1-\beta_{1})}  + \frac{\beta_{1} \rho_{a}D_{\Delta}}{2(1-\eta_{1}\eta_{a})(1-\beta_{1})} + \frac{\rho_{a}\sqrt{1-\beta_{2}}D_{u,1} D_{\Delta}}{2\alpha_{1}^{-}(1-\sqrt{\beta_2})(1-\eta_{a})^2(1-\beta_{1})}\notag\\
				&  + \frac{\rho_{a}^2 D_{\Delta}}{2(1-\eta_{a}^2)(1-\beta_{1})} + \frac{3\rho_a^2  D_{\phi} D_{b,2}}{2(1-\eta_a^2)(1-\beta_{1})}+\frac{3D_{u,1}^2 D_{\phi}}{(1-\eta_1^2)(1-\beta_1)^3}+\frac{3\rho_{a}^2D_{b,1}^2D_{\phi}}{(1-\eta_{a}^2)(1-\beta_1)} +\frac{D_{\Delta_{u,2}}D_{\phi}}{1-\beta_1} \label{regbound}\\
				&+\left[\frac{ D_{u,1}}{(1-\beta_1)(1-\eta_1)^2}+  \frac{\rho_{a}D_{b,1}}{(1-\eta_a)^2}\right]\frac{\sqrt{1-\beta_{2}}D_{u,1} D_{\Delta}}{2\alpha_{1}^{-}(1-\sqrt{\beta_2})(1-\beta_{1})}+  \left[\frac{ D_{u,1}}{(1-\beta_1)(1-\eta_1)(1-\eta_{a})}+  \frac{\rho_{a}D_{b,1}}{(1-\eta_a)^2}\right]\frac{D_{\Delta}  \rho_{a}}{2(1-\beta_{1})}.\notag
			\end{align}
			\hrulefill
			\vspace*{4pt}
		\end{figure*}

		\subsection{Transmit Power Allocation for Multiple Targets}

		Given $\{{\mathbf{u}}_q^{(k,j+1)}\}_{q=1}^{Q}$, the problem in  (\ref{subj1}) can be expressed as
		\begin{equation}\label{prob0p}
			\begin{split}
				\min_{\mathbf{p}^{(k)}\in\mathcal{S}_p} \;&\sum_{q=1}^Q \mathcal{F}_{\mathrm{pa}}(p_q^{(k)}),\\
			\end{split}
		\end{equation}
		where $\mathcal{F}_{\mathrm{pa}}(p_q^{(k)})=\log \det \mathbf{C}_q(p_q^{(k)}|\mathbf{u}_q^{(k,j)})$ is the cost function and $\mathcal{S}_p=\{\mathbf{p}^{(k)}|\sum_{q=1}^Q p_q^{(k)} \leq P_T,p_q^{(k)} \geq P_{\min}, q=1,2\cdots, Q\}$ denotes the feasible set of $\mathbf{p}^{(k)}$.
		This problem is convex and can be reformulated as a SDP problem, i.e.,
		\begin{equation}\label{prob0pasdp}
			\begin{split}
				\max_{\mathbf{p}^{(k)}} \;&\sum_{q=1}^Q \log \det(\mathbf{Q}_q),\\
				\text{s.t.} \; &\sum_{q=1}^Q p_q^{(k)} \leq P_T,\quad p_q^{(k)} \geq P_{\min}, \\
				&\mathbf{J}_q^{(k)}(p_q^{(k)}|\mathbf{u}_q^{(k,j)}) \succeq \mathbf{Q}_q, q=1,2\cdots, Q,
			\end{split}
		\end{equation}
		where $\{\mathbf{Q}_q\}_{q=1}^Q$ denotes a set of auxiliary symmetric matrices. Then, this problem can be solved by the CVX toolbox. 
		
		However, the CVX toolbox is generally time-consuming, especially when the number of targets is large.
		To reduce the computational complexity and reveal more physical insights, 
		we propose an iterative water-filling-based power allocation method. 
		First, we merge the total power constraint into the cost function by the Lagrange multiplier method, i.e.,
		\begin{equation}\label{Lag11}
			\begin{split} 
				\mathcal{L}_{\mathrm{pa}}(\mathbf{p}^{(k)})&=\sum_{q=1}^Q \mathcal{F}_{\mathrm{pa}}(p_q^{(k)})
				+ \lambda_{\mathrm{pa}}(P_T-\sum_{q=1}^Q p_q^{(k)}),
			\end{split}
		\end{equation}
		where $\lambda_{\mathrm{pa}}$ is the Lagrange multiplier.
		The derivative of (\ref{Lag11}) w.r.t. $p_q^{(k)}$ is given by
		\begin{equation}\label{Lag11diff}
			\begin{split} 
				\frac{\partial \mathcal{L}_{\mathrm{pa}}(\mathbf{p}^{(k)})}{\partial p_q^{(k)}}=\tr( (\mathbf{J}_{P,q}^{(k)}+p_q^{(k)}\widetilde{\mathbf{\Sigma}}_{q}^{(k)})^{-1}\widetilde{\mathbf{\Sigma}}_{q}^{(k)})- \lambda_{\mathrm{pa}},
			\end{split}
		\end{equation}
		where $	\widetilde{\mathbf{\Sigma}}_{q}^{(k)}=\sum_{n=1}^{N} u_{q,n}^{(k)}\overline{\mathbf{M}}_{q,n}^{(k)}.$
		By setting $\frac{\partial \mathcal{L}_{\mathrm{pa}}(\mathbf{p}^{(k)})}{\partial p_q^{(k)}}=0$, we have the following fixed-point equation, i.e.,
		\begin{equation}\label{wffixpoint}
			\begin{split} 
				p_q^{(k)}=\frac{1}{\lambda_{\mathrm{pa}}}-\frac{\tr\;(\mathbf{J}_{P,q}^{(k)}+p_q^{(k)}\widetilde{\mathbf{\Sigma}}_{q}^{(k)})^{-1}\mathbf{J}_{P,q}^{(k)}}{\tr\;(\mathbf{J}_{P,q}^{(k)}+p_q^{(k)}\widetilde{\mathbf{\Sigma}}_{q}^{(k)})^{-1}\widetilde{\mathbf{\Sigma}}_{q}^{(k)}}.
			\end{split}
		\end{equation}
		If $\mathbf{J}_{P,q}^{(k)}$ and $\widetilde{\mathbf{\Sigma}}_{q}^{(k)}$ reduce to one-dimensional constants denoted by ${J}_{P,q}^{(k)}$ and  $\widetilde{{\Sigma}}_{q}^{(k)}$, respectively, the closed-form solution of $p_q^{(k)}$ can be directly obtained from (\ref{wffixpoint}), i.e., $p_q^{(k)}=\mu_{\mathrm{wf}}-{{J}_{P,q}^{(k)}}/{\widetilde{{\Sigma}}_{q}^{(k)}}$, 
		where $\mu_{\mathrm{wf}}=\frac{1}{\lambda_{\mathrm{pa}}}$ denotes the water level. For the matrix-version $\mathbf{J}_{P,q}^{(k)}$ and $\widetilde{\mathbf{\Sigma}}_{q}^{(k)}$, 
		we propose to obtain $p_q^{(k)}$ and the water level $\mu_{\mathrm{wf}}$ by an iteration process. In particular, at the $i$th iteration, $p_{q,i+1}^{(k)}$ is obtained by 
		\begin{equation}
			p_{q,i+1}^{(k)}=\left\lfloor \mu_{\mathrm{wf}}-
			\frac{\tr\;(\mathbf{J}_{P,q}^{(k)}+p_{q,i}^{(k)}\widetilde{\mathbf{\Sigma}}_{q}^{(k)})^{-1}\mathbf{J}_{P,q}^{(k)}}{\tr\;(\mathbf{J}_{P,q}^{(k)}+p_{q,i}^{(k)}\widetilde{\mathbf{\Sigma}}_{q}^{(k)})^{-1}\widetilde{\mathbf{\Sigma}}_{q}^{(k)}}\right\rfloor_{P_{\min}},
		\end{equation}
		where $p_{q,i}^{(k)}$ denotes the power for the $q$th target at the $i$th iteration and $\left\lfloor a\right\rfloor_{b}=\max\{a,b\}$. Then, the water level
		$\mu_{\mathrm{wf}}$ is updated by setting $\sum_{q=1}^Q p_{q,i+1}^{(k)}(\mu_{\mathrm{wf}}) = P_T.$

		\begin{remark}
			According to the Rayleigh quotient, we have
			$\tilde{\lambda}_{\min}\leq
			\frac{\tr\;(\mathbf{J}_{P,q}^{(k)}+p_q^{(k)}\widetilde{\mathbf{\Sigma}}_{q}^{(k)})^{-1}\mathbf{J}_{P,q}^{(k)}}{\tr\;(\mathbf{J}_{P,q}^{(k)}+p_q^{(k)}\widetilde{\mathbf{\Sigma}}_{q}^{(k)})^{-1}\widetilde{\mathbf{\Sigma}}_{q}^{(k)}}
			\leq\tilde{\lambda}_{\max}$, 
			where $\tilde{\lambda}_{\min}$ and $\tilde{\lambda}_{\max}$ denote the  minimum and maximum eigenvalue of $(\widetilde{\mathbf{\Sigma}}_{q}^{(k)})^{-1}\mathbf{J}_{P,q}^{(k)}$, respectively. Note that $\mathbf{J}_{P,q}^{(k)}$ and $\widetilde{\mathbf{\Sigma}}_{q}^{(k)}$ denote the FIM of the prediction and the measurement, respectively. Thus, the eigenvalues of $(\widetilde{\mathbf{\Sigma}}_{q}^{(k)})^{-1}\mathbf{J}_{P,q}^{(k)}$ denote the ratio between the prediction and measurement. Recalling (\ref{wffixpoint}), if the eigenvalues of $(\widetilde{\mathbf{\Sigma}}_{q}^{(k)})^{-1}\mathbf{J}_{P,q}^{(k)}$ are larger,  $p_q^{(k)}$ will be lower. This indicates that, more power will be allocated to a target, if 1) the measurement provides more information than the prediction, which enables the system to improve the accuracy of the prediction, or 2) the prediction of this target is so bad such that the system needs to allocate more power for better motion state estimation. 
			In turn, if the eigenvalues of $(\widetilde{\mathbf{\Sigma}}_{q}^{(k)})^{-1}\mathbf{J}_{P,q}^{(k)}$ are smaller,  $p_q^{(k)}$ will be lower. This indicates that, a target will be assigned with a lower power, if 1) the prediction is good enough; or 2) the measurement is too bad.
		\end{remark}



		\section{Simulation}
		In the simulation, we will show the efficiency and effectiveness of the proposed DAN and FP-WF algorithms. In the following, we first introduce the system parameters, the training details of DAN, and the benchmark algorithms.
		
		\textbf{System parameters}:
		We consider a mmWave system operating at a carrier frequency of 28 GHz.  There is one BS acting as the transmitter, which is located at $[0,0]$ m. The number of SNs is $N=32$. These SNs are uniformly distributed in the area within $400\times 400$ $\text{m}^2$. On average, there is one SN within an area of 5000 $\text{m}^2$.  The measurement covariance defined in (\ref{meacov}) is generated by
		$\mathbf{\Sigma}_{q,n}^{(k)}=\frac{1}{\mathrm{SNR}_{q}^{(k)}}\dot{\mathbf{\Sigma}}_{q,n}^{(k)}$, where $\dot{\mathbf{\Sigma}}_{q,n}^{(k)}=\diag[\dot{\sigma}_{\theta_{q,n}^{(k)}}^2,\dot{\sigma}_{\tau_{q,n}^{(k)}}^2,\dot{\sigma}_{\mu_{q,n}^{(k)}}^2]$ with  $\dot{\sigma}_{\theta_{q,n}^{(k)}}=2$, $\dot{\sigma}_{\tau_{q,n}^{(k)}}=1$, $\dot{\sigma}_{\mu_{q,n}^{(k)}}=1$. 
		The SNR is defined by $\mathrm{SNR}_{q}^{(k)} =\frac{p_q^{(k)}\gamma_0}{\sigma^2(d_{q,n}^{(k)})^2},$ where $\gamma_0=-61.4$ dB denotes the pathloss at reference distance. 
		We set the total power at BS $P=30$ dBm, the minimum power for single target $P_{min}=20$ dBm, the noise power $\sigma^2=-90$ dBm, the intensity of process noise $q_s=5$, and $\Delta T =0.5$ s.


		\textbf{Initialization of motion state}:
		There are three targets to be tracked, i.e.,  $Q=3$, if not otherwise specified.
		The initial velocities of the targets are given as $\mathbf{v}_{1}=[-10,0]^\TT$ m/s, $\mathbf{v}_{2}=[0,-10]^\TT$ m/s, $\mathbf{v}_{3}=[10,0]^\TT$ m/s, respectively. The initial locations of the targets are given as $\mathbf{x}_{1}^{(0)}=[124, 124]^\TT$ m, $\mathbf{x}_{2}^{(0)}=[-134, 134]^\TT$ m, and $\mathbf{x}_{3}^{(0)}=[-144, -144]^\TT$ m, respectively.

		\textbf{Training details}:
		During training, the learnable parameters are optimized by the SGD optimizer in the PyTorch with a learning rate $5\times10^{-5}$.  
		In our experiment, the loss function for training is selected as
		$f_{loss}=\frac{1}{L}\sum_{l=1}^L ||\mathbf{u}_{ES} - \hat{\mathbf{u}}^{l}||^2,$ where $\mathbf{u}_{ES}$ denotes the selection vector obtained by the exhaustive search (ES). The number of data for training is set as $N_{\text{train}}=500$. 
		The network parameters are set as $\rho=1$, $\rho_{a}=10^2$, $\gamma=10^4$, $\beta_{2}=0.999$, $\eta_{1}=0.99$, and $\eta_{a}=0.99$. The learnable parameters are initialized as  $\beta_{1}=0.99$,  and $\alpha_1 = 0.15$ for all layers. The number of layers is set as $L=10$. The maximum number of ADMM iterations is set as $200$. 

		%


		\textbf{Benchmark methods}: The proposed methods are compared with the following algorithms for SN selection and power allocation. 
		
		\emph{1) SN selection:}
		We compare DAN with the following  methods:
		
		$\bullet$ `Nearest SN Selection': this method selects the subset of SNs nearest to the target;
		
		$\bullet$ `Exhaustive Search (ES)': this method selects the subset of SNs which minimizes the cost function;
		
		$\bullet$ `MM-CVX': the method solves the optimization problem (\ref{prob1Gu}) by CVX toolbox. 
		
		$\bullet$ `MM-ADMM': the optimization-based method proposed in Sec. III. A. To show the impact of $\mathbf{T}^{(k,l)}$ in MM-ADMM, we use two different $\mathbf{T}^{(k,l)}$. Specifically, the first choice is  $\mathbf{T}_1^{(k,l)}=\tr(\mathbf{H}_F(\mathbf{u}^{(k,l)}))\mathbf{I}$, and the second choice is $\mathbf{T}_2^{(k,l)}=\lambda_{\max}(\mathbf{H}_F(\mathbf{u}^{(k,l)}))\mathbf{I}$, which are denoted by `MA-I' and `MA-II', respectively. The parameters of MM-ADMM and MM-CVX are the same as that for DAN. 
		The maximum number of MM iterations for MM-ADMM and MM-CVX is set as $30$ and $50$, respectively, unless specified otherwise.
		
		\emph{2) Power allocation:}
		We compare FP-WF with `CVX', which represents the method for solving (\ref{prob0pasdp}) by CVX. 
		
		\subsection{Computational Cost}
		\begin{table}[!t]
			\caption{Running time in second for joint SN selection (ES, MA-I, MA-II, DAN) and power allocation (CVX, FP-WF) (Averaged Over 1000 Monte-Carlo trials) \label{Time:overall}}
			\centering
			\begin{tabular}{c|c c c c c}
				\toprule[1.5pt]
				Method & ES & MM-CVX & MA-I & MA-II & DAN\\
				\midrule[1pt]
				CVX & 18.6242 & 24.4350 & 10.0120	& 13.7984 & 6.6404\\
				FP-WF & 13.5733 & 19.4870 & 4.5109 & 9.1722 & 0.7724\\
				\toprule[1.5pt]
			\end{tabular}
		\end{table}

		Table \ref{Time:overall} shows the running time\footnote{Configuration of this computer: CPU: Inter Core i9-9900 @3.10GHz; RAM: 16GB; Software: Python 3.10.9 in Microsoft visual studio code and Matlab 2020b.} of the algorithms composed of different power allocation and SN selection methods.	
		It can be observed that the running time of  DAN $\&$ FP-WF is 0.7724 s, which is the lowest among all combinations. 
		Meanwhile, we can observe that the running time of ES $\&$ CVX  is 18.6242 s, which is about 24.11 times more than that of DAN $\&$ FP-WF. 
		To further demonstrate the low computational complexity provided by DAN and FP-WF, we study the computational cost of the SN selection and power allocation methods, respectively.

		\begin{table}[!t]
			\caption{Running time in second for SN selection (ES, MA-I, MA-II, DAN) (Averaged Over 1000 Monte-Carlo trials) \label{tab:SNs}}
			\centering
			\begin{tabular}{c|c c c c c}
				\toprule[1.5pt]
				$N$ & ES & MM-CVX & MA-I & MA-II & DAN \\
				\midrule[1pt]
				$32$  & 4.2662 & 6.4081  & 1.3712  & 2.8322 & 0.2453\\
				$64$  & 34.7366 & 23.0416 &  3.9726 & 9.1618 & 0.3263 \\
				$128$  & 280.1406 & 93.2871  & 18.9184 & 40.0574 & 0.6316\\
				\toprule[1.5pt]
			\end{tabular}
		\end{table}
		\textbf{Running time of the SN selection methods}: Table \ref{tab:SNs} shows the running time of the SN selection algorithms with different $N$. 
		DAN achieves the lowest computational cost among the candidates with different $N$. 
		The computational consumption of ES is extremely large, especially when $N$ is large. 
		For example, when $N=128$, the DAN is about 443 times faster than ES.  MM-CVX is more time-consuming than MM-ADMM.
		Meanwhile, the running time of DAN is less than that of the MM-ADMM. 
		There are two main reasons: 1) one layer of DAN has a lower computational cost than one iteration of MM-ADMM. In particular, DAN only requires the gradient, while MM-ADMM requires both the gradient and Hessian matrix, which needs more computational cost, and 2) owing to the well-trained $\mathbf{T}^{(k,l)}$, DAN can converge faster than MM-ADMM, which will be shown in the following.

\begin{figure}[t]
			\centering
			\includegraphics[width=3.42in]{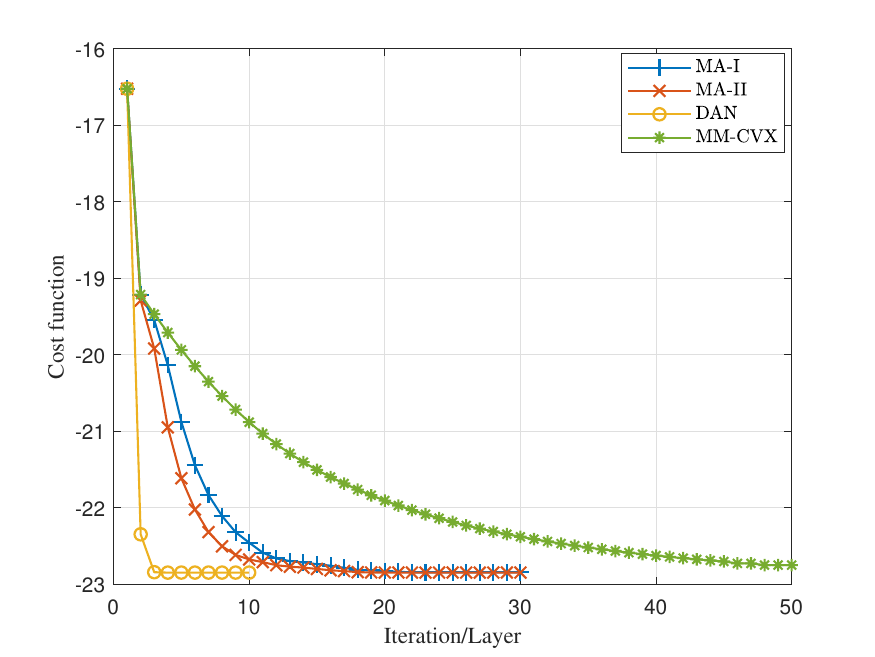}\
			\caption{Cost function over the number of iterations/layers.}
			\label{fig_Converge}
		\end{figure}

		\textbf{Convergence of the SN selection methods}: 
		The running time of MM-CVX, MM-ADMM and DAN	is proportional to the required number of  iterations/layers to converge.
		Fig. \ref{fig_Converge} shows the cost function over the number of the iterations (optimization-based methods) or the layers (DAN). 
		First, MM-CVX needs about 50 iterations to converge, which is more than MM-ADMM and DAN. 
		Meanwhile, we can observe that DAN can converge within 3 layers, while MM-ADMM needs about 15-20 iterations to converge, which leads to more running time.
		This is because, 
		unlike MM-ADMM, DAN utilizes the momentum, which accumulates the gradient of the past layers and can thus speed up the convergence \cite{kingma2014adam}. 
		Meanwhile, we see that MM-ADMM-II can converge faster than MM-ADMM-I which indicates that the convergence of MM-ADMM highly depends on the choice of $\mathbf{T}^{(k,l)}$. This is also the motivation to learn $\mathbf{T}^{(k,l)}$ in DAN.
		
		\textbf{Running time of the power allocation methods}:
		Table \ref{tab:PA} shows the running time for the power allocation algorithms versus different $Q$. We can observe that the running time of FP-WF is much lower than CVX for different cases. This is because FP-WF is derived based on the Lagrange multiplier method, which can solve (\ref{prob0pasdp}) more efficiently than the interior point method used by CVX.
		
		\begin{table}[!t]
			\caption{Running time in second for power allocation (CVX, FP-WF) (Averaged Over 1000 Monte-Carlo trials) \label{tab:PA}}
			\centering
			\begin{tabular}{c|c c c c}
				\toprule[1.5pt]
				Method & $Q=3$ & $Q=4$& $Q=5$& $Q=6$\\
				\midrule[1pt]
				CVX & 1.7246 & 1.9854 & 2.0828 & 2.2966\\
				FP-WF & 0.0030 & 0.0060 & 0.0072 & 0.0087\\
				\toprule[1.5pt]
			\end{tabular}
		\end{table}
		
		\subsection{Tracking Accuracy}
\begin{figure}[t]
			\centering
			\includegraphics[width=3.42in]{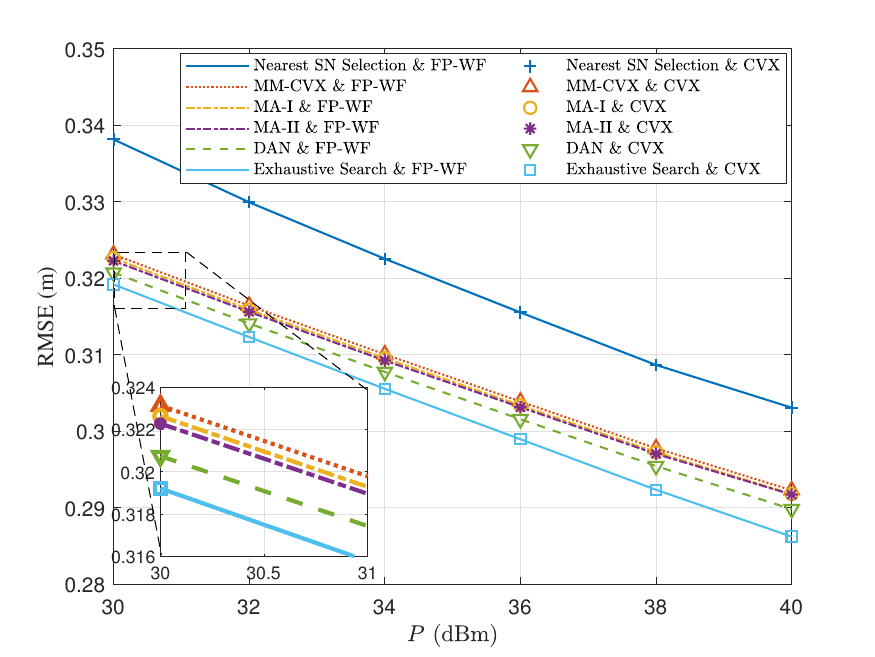}\
			\caption{Average RMSE versus the total power budget $P$.}
			\label{fig_AO}
		\end{figure}

		The average root mean square error (RMSE) of multiple targets tracking over $Q$ targets and $K$ frames is selected as the performance metric for multiple target tracking, which is defined as  $\frac{1}{Q}\frac{1}{K}\sum_{q=1}^{Q}\sum_{k=1}^{K}\sqrt{\frac{1}{N_{mc}}\sum_{i=1}^{N_{mc}}\Vert\mathbf{x}_q^{(k)}-\hat{\mathbf{x}}_q^{(k,i)} \Vert^2}$, where 
		$\hat{\mathbf{x}}_q^{(k,i)}$ denotes the estimated position of the target $q$ at the $k$th time frame in the $i$th Monte-Carlo trial, and $N_{mc}$ denotes the number of Monte-Carlo trials. 
		The number of tracking frames is set as $K=10$. Fig. \ref{fig_AO} shows the average RMSE with different power budget $P$.
		We have several observations. First, associated with different SN selection methods, FP-WF can achieve the same performance as CVX. Recalling from the results in Table \ref{tab:PA}, compared to CVX, FP-WF can reduce the computational cost without any performance loss.			
		Second, we can observe that ES can achieve the best performance among the SN selection methods. However, from Table \ref{Time:overall}, it can be observed that the running time of ES is extremely high, which limits its real application. 
		Third, MM-CVX and MM-ADMM can achieve similar performance, but as shown in Table \ref{Time:overall},	the computational cost of MM-CVX is higher than that of MM-ADMM.
		Furthermore, DAN can outperform MM-ADMM, which is because a more suitable $\mathbf{T}$ is learned by DAN. 
		Finally, the performance of the nearest SN selection is worse than DAN. This is because the tracking performance is affected by both the distance and the angle from target to SNs. DAN takes both of them into consideration, while the nearest SN selection only considers the distance. 
		This will be further demonstrated in the next part. 

		\begin{figure}[t]
			\centering
			\subfloat[]{\includegraphics[width=1.71in]{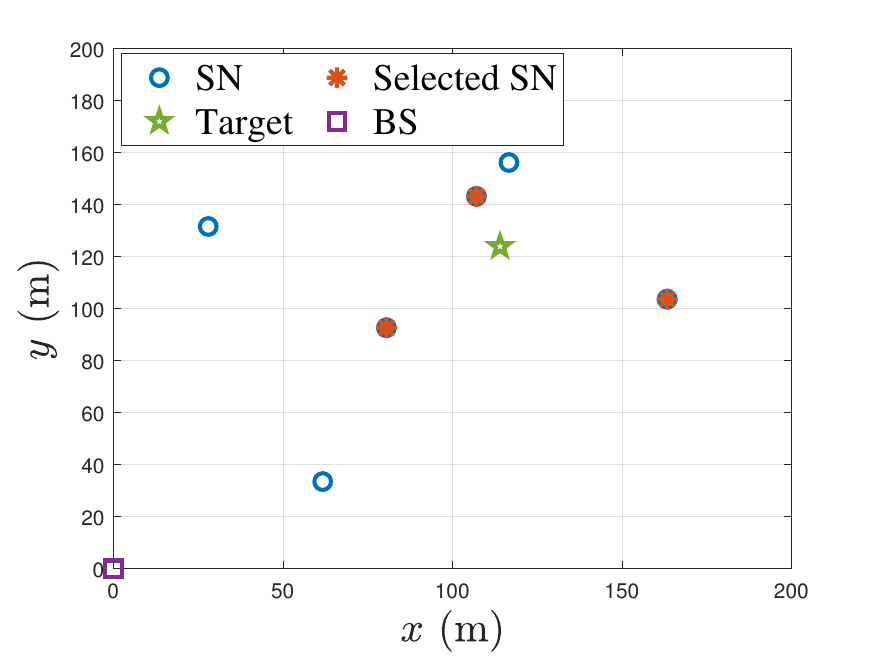}}\
			\subfloat[]{\includegraphics[width=1.71in]{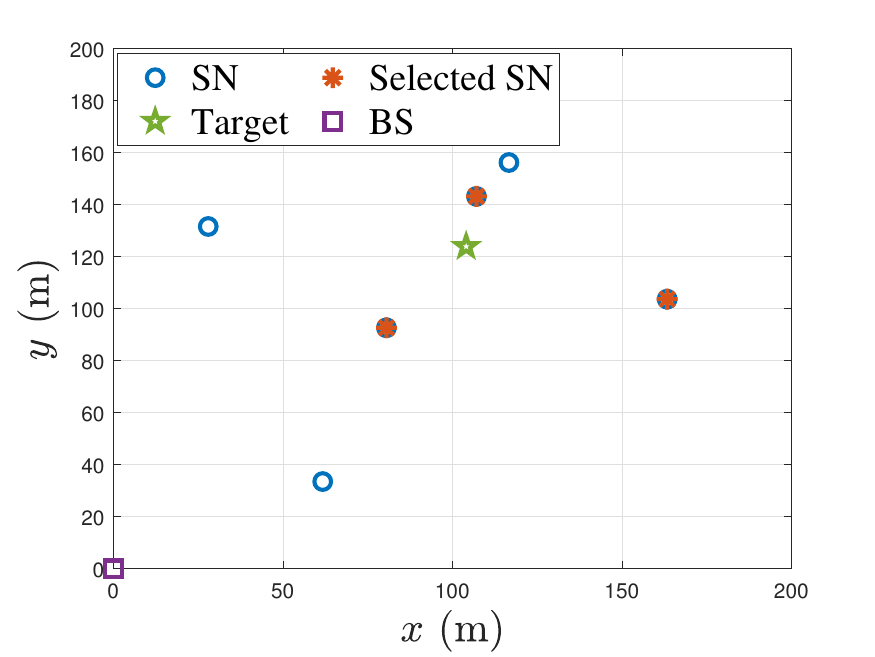}}\\
			\subfloat[]{\includegraphics[width=1.71in]{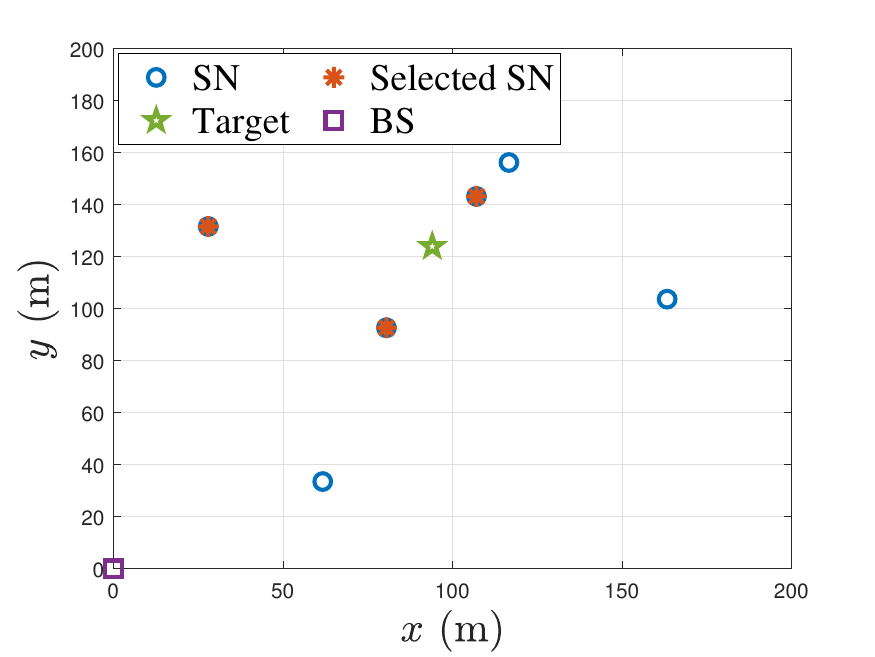}}\
			\subfloat[]{\includegraphics[width=1.71in]{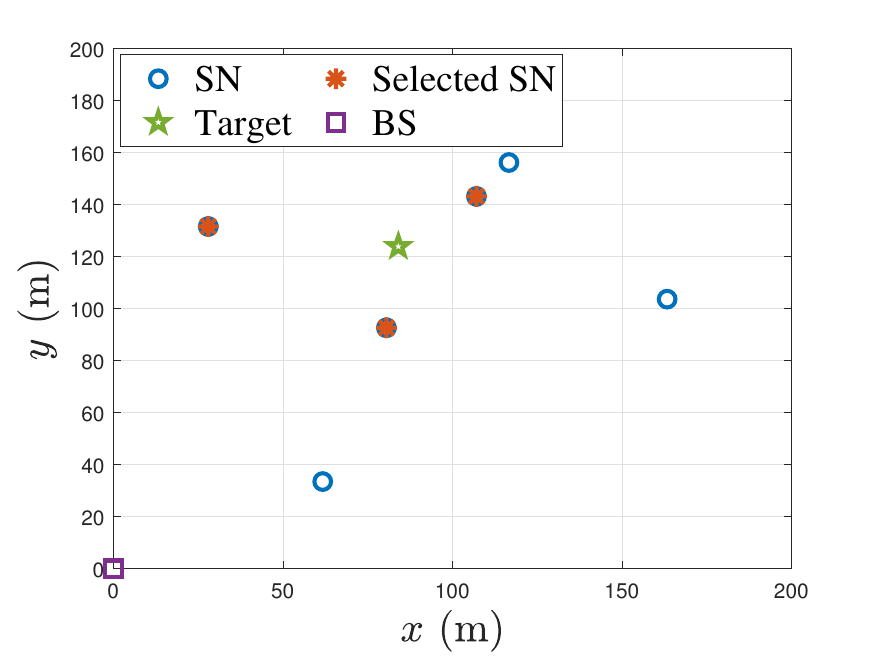}}\\
			\caption{SN selection result by DAN at 4 consecutive frames. (a) Frame 2; (b) Frame 4; (c) Frame 6; (d) Frame 8;
			}
			\label{fig_map_frame}
		\end{figure}
		
		\begin{figure}[t]
			\centering
			\includegraphics[width=3.42in]{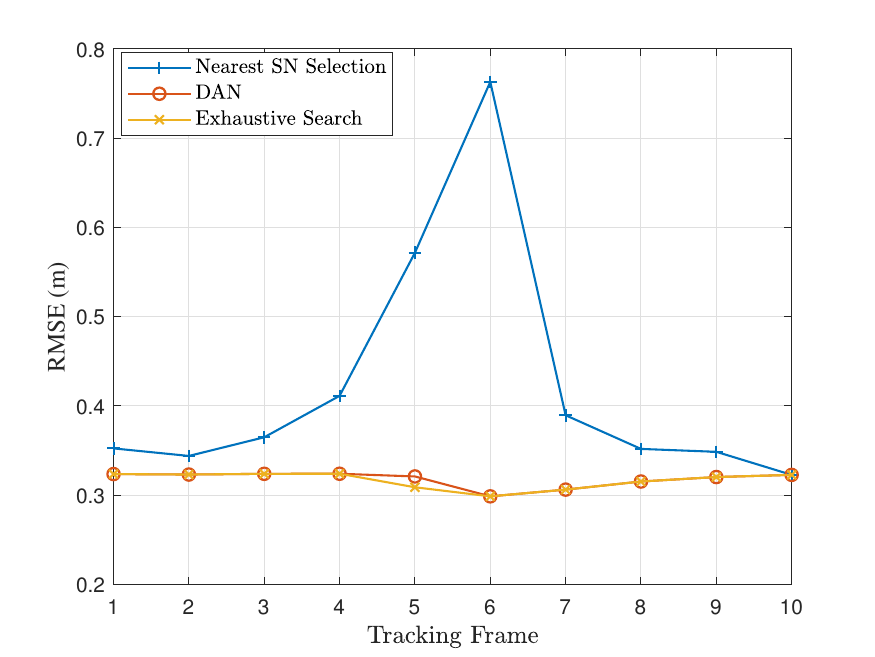}
			\caption{RMSE over tracking frames.
			}
			\label{fig_rmse_fix}
		\end{figure}


		
		\textbf{Illustration of SN selection}: 
		To better understand the effect of SN selection, we focus on the single target case in this section. The power allocated to the target is set as $p=25$ dBm. The initial state of the target is given by
		$\mathbf{v}=[-10,0]^\TT$ m/s and 	$\mathbf{x}^{(0)}=[124, 124]^\TT$ m. 
		Fig. \ref{fig_map_frame} shows the SN selection result by DAN in 4 consecutive frames. 
		The selection depends on the geometric relation between the target and SNs. DAN does not always choose the nearest SNs, because, besides the distance, the different perspectives to observe the target provided by different SNs will also affect the tracking performance. Fig. \ref{fig_rmse_fix} shows the corresponding RMSE over the tracking frames. 
		It can be observed that DAN consistently outperforms the Nearest SN selection and achieves comparable performance as ES. 
		


		\textbf{Effect of noise power}: 
		\begin{figure}[t]
			\centering
			\includegraphics[width=3.42in]{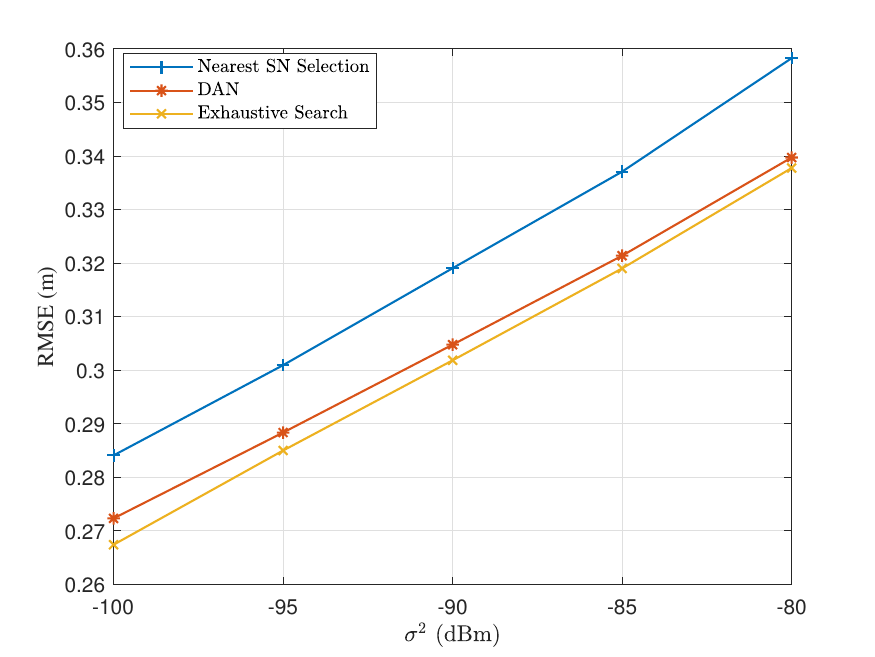}\
			\caption{RMSE versus different noise power $\sigma^2$.}
			\label{fig_RMSE_diffsigma2}
		\end{figure}
		One of the biggest drawbacks of DL-based approaches is the performance degradation when the features (such as the noise power) in test data differ from those in training. This leads to the study of generalization in this part.
		Fig. \ref{fig_RMSE_diffsigma2} shows the performance under different noise power with $N=32$. When the noise power is different from that of the training data, DAN can provide a near-ES RMSE. It indicates that DAN can adapt to the change of $\sigma^2$, which makes DAN attractive in real applications.

		\subsection{Accuracy-Complexity Tradeoff}
\begin{figure}[t]
			\centering
			\includegraphics[width=3.42in]{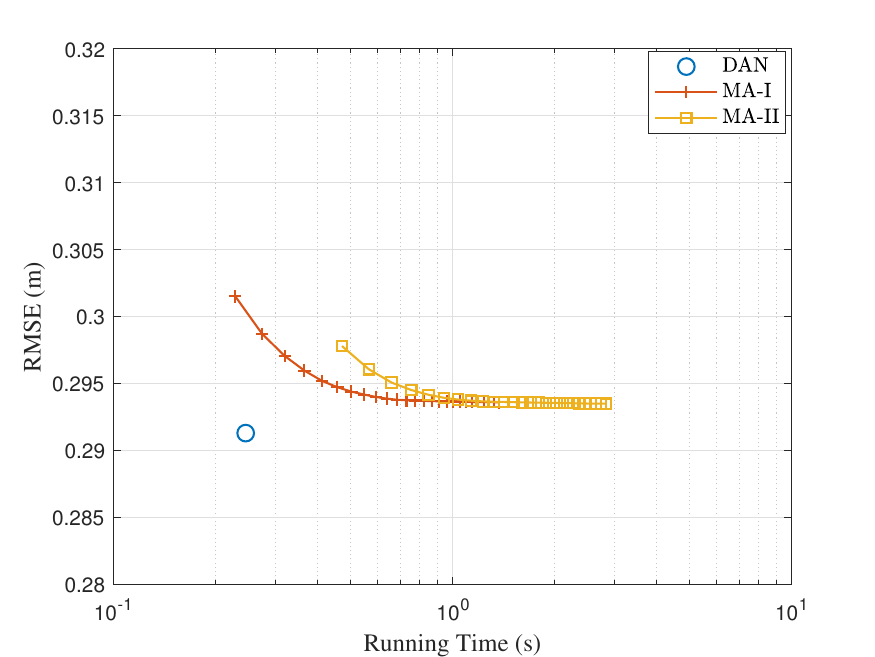}\
			\caption{NMSE versus the running time.}
			\label{fig_rt}
		\end{figure}

		By adjusting the termination tolerance and the maximum number of iterations, a tradeoff between computational cost and accuracy can be achieved by MM-ADMM. Meanwhile, the proposed DAN requires a fixed number of layers and thus has a fixed running time. Fig. \ref{fig_rt} shows the RMSE performance of different algorithms versus the running time. It is observed that DAN can always outperform MM-ADMM in terms of both computational cost and RMSE. Moreover, though MM-ADMM-II can converge faster than MM-ADMM-I, $\mathbf{T}_2^{(k,l)}$ requires more computational cost than $\mathbf{T}_1^{(k,l)}$. Thus, given the same time cost,  MM-ADMM-I outperforms MM-ADMM-II.

		\section{Conclusion}
		In this paper, we considered the joint SN selection and power allocation problem for tracking multiple maneuvering targets in PMNs. To meet the stringent latency requirement of sensing applications, we proposed a model-driven DL-based approach for SN selection by unfolding the optimization-based MM-ADMM method.
		A novel DNN architecture was derived to speed up the convergence by exploiting the momentum, where the convergence property was guaranteed by deriving the regret bound. Furthermore, we proposed an efficient power allocation method based on fixed-point water filling and revealed some physical insights. Simulation results demonstrated that the proposed method can achieve better performance than existing optimization-based methods with much lower computational cost. This work demonstrated that, by reducing the number of iterations and improving the effectiveness of each layer, model-driven DL-based approaches offer a promising solution to meet the stringent latency requirement of sensing applications. 
		


		\appendices
		
		\section{Proof of Theorem \ref{MMconvergence}}
		\label{MMconvergenceProof}
		Given $\mathcal{L}(\mathbf{u}^{(k)})$ is convex,
		we have
		\begin{equation}\label{profinal}
			\begin{split}
				\mathcal{G}(\mathbf{u}^{(k,l)})-\mathcal{G}(\mathbf{u}^{(k,\star)})\leq  \left<\mathbf{d}_{u}^{(k,l)}, \Delta\mathbf{u}^{(k,l)}\right>,
			\end{split}
		\end{equation}
		where $\Delta \mathbf{u}^{(k,l)}=\mathbf{u}^{(k,l)} - \mathbf{u}^{(k,\star)}$. 
		Since $R_L\leq \sum_{l=1}^L\left<\mathbf{d}_{u}^{(k,l)}, \Delta\mathbf{u}^{(k,l)}\right>$, the main idea of the proof is to find an upperbound of $\sum_{l=1}^L\left<\mathbf{d}_{u}^{(k,l)}, \Delta\mathbf{u}^{(k,l)}\right>$.
		Recalling from (\ref{um2netstar}), we have
		\begin{equation}\label{eq50}
			\begin{split}
				&\Vert \mathbf{\Phi}_{l}^{\frac{1}{2}}\Delta \mathbf{u}^{(k,l+1)}\Vert^2=\Vert \mathbf{\Phi}_{l}^{\frac{1}{2}}(\mathbf{u}^{(k,l+1)} - \mathbf{u}^{(k,\star)})\Vert^2\\
				&\overset{(a)}{=}\Vert\mathbf{\Phi}_{l}^{\frac{1}{2}}(\mathbf{u}^{(k,l)}-\mathbf{\Phi}_{l}^{-1}(\mathbf{d}_{\star}^{(k,l)}-\nu_{l}\mathbf{1}))-\mathbf{u}^{(k,\star)}\Vert^2\\
				&\overset{(b)}{=}\Vert\mathbf{\Phi}_{l}^{\frac{1}{2}}\Delta \mathbf{u}^{(k,l)}-\mathbf{\Phi}_{l}^{-\frac{1}{2}}(\hat{\mathbf{m}}_{l}-\rho_{a,l}\hat{\mathbf{b}}_{l}-\nu_{l}\mathbf{1})\Vert^2\\
				&\overset{(c)}{=}\left\Vert \mathbf{\Phi}_{l}^{\frac{1}{2}}\Delta \mathbf{u}^{(k,l)} \right\Vert^2 - 2 \left<(1-\beta_{1,l})\mathbf{d}_{u}^{(k,l)},\Delta \mathbf{u}^{(k,l)}\right>\\
				& \quad - 2 \left<\beta_{1,l}\hat{\mathbf{m}}_{l-1}-\rho_{a,l}\hat{\mathbf{b}}_{l}-\nu_{l}\mathbf{1},\Delta \mathbf{u}^{(k,l)}\right>\\
				&\quad+ \Vert \mathbf{\Phi}_{l}^{-\frac{1}{2}}(\hat{\mathbf{m}}_{l}
				-\rho_{a,l}\hat{\mathbf{b}}_{l}-\nu_{l}\mathbf{1}
				) \Vert^2,
			\end{split}
		\end{equation}
		where step (a) follows (\ref{um2netstar}), step (b) follows (\ref{dstar}), 
		and step (c) follows  (\ref{firstordermom}).
		
		By adding $2 \left<(1-\beta_{1,l})\mathbf{d}_{u}^{(k,l)},\Delta \mathbf{u}^{(k,l)}\right>-\Vert \mathbf{\Phi}_{l}^{\frac{1}{2}}\Delta \mathbf{u}^{(k,l+1)}\Vert^2$ to both sides of (\ref{eq50}), and dividing both sides of (\ref{eq50}) by $2(1-\beta_{1,l})$, we have
		
		
		\begin{align}
			&\left<\mathbf{d}_{u}^{(k,l)},\Delta \mathbf{u}^{(k,l)}\right> =\frac{\Vert \mathbf{\Phi}_{l}^{\frac{1}{2}}\Delta \mathbf{u}^{(k,l)}\Vert^2}{2(1-\beta_{1,l})} -\frac{\Vert \mathbf{\Phi}_{l}^{\frac{1}{2}}\Delta \mathbf{u}^{(k,l+1)}\Vert^2}{2(1-\beta_{1,l})}\notag\\
			&= -\frac{\left<\beta_{1,l}\hat{\mathbf{m}}_{l-1},\Delta \mathbf{u}^{(k,l)}\right>}{1-\beta_{1,l}}+\frac{\left<\rho_{a,l} \hat{\mathbf{b}}_{l},\Delta \mathbf{u}^{(k,l)}\right>}{1-\beta_{1,l}} \label{grieq00}\\
			&\quad +\frac{\left<\nu_{l}\mathbf{1},\Delta \mathbf{u}^{(k,l)}\right>}{1-\beta_{1,l}}+\frac{\Vert \mathbf{\Phi}_{l}^{-\frac{1}{2}}(\hat{\mathbf{m}}_{l}-\rho_{a,l}\hat{\mathbf{b}}_{l}-\nu_{l}\mathbf{1}) \Vert^2}{2(1-\beta_{1,l})}.\notag
		\end{align}

		By using the Young's inequality for products, i.e., $\pm ab\leq \frac{a^2}{2} +\frac{b^2}{2}$, the second, third, and fourth terms on the right-hand side of (\ref{grieq00}) are upperbounded by $-\frac{\left<\beta_{1,l}\hat{\mathbf{m}}_{l-1},\Delta \mathbf{u}^{(k,l)}\right> }{1-\beta_{1,l}}\leq \frac{ \Vert \mathbf{\Phi}_{l}^{-\frac{1}{2}}\hat{\mathbf{m}}_{l-1}	\Vert^2 }{2(1-\beta_{1})} +\frac{\Vert \mathbf{\Phi}_{l}^{\frac{1}{2}}\Delta \mathbf{u}^{(k,l)}\Vert^2}{2(1-\beta_{1})}$, 
		$\frac{\left< \hat{\mathbf{b}}_{l},\Delta \mathbf{u}^{(k,l)}\right>}{1-\beta_{1,l}}\leq\frac{\Vert \mathbf{\Phi}_{l}^{-\frac{1}{2}}\hat{\mathbf{b}}_{l}\Vert^2}{2(1-\beta_{1})} +  \frac{\Vert \mathbf{\Phi}_{l}^{\frac{1}{2}}\Delta \mathbf{u}^{(k,l)}\Vert^2}{2(1-\beta_{1})}$, and 
		$\frac{\left<\mathbf{1},\Delta \mathbf{u}^{(k,l)}\right>}{1-\beta_{1,l}}\leq  \frac{\Vert \mathbf{\Phi}_{l}^{-\frac{1}{2}}\mathbf{1}\Vert^2}{2(1-\beta_{1})} + \frac{\Vert \mathbf{\Phi}_{l}^{\frac{1}{2}}\Delta \mathbf{u}^{(k,l)}\Vert^2}{2(1-\beta_{1})}$, respectively.
		By utilizing the inequality between the arithmetic mean and quadratic mean, the last term on the right-hand side of (\ref{grieq00}) is upperbounded by $\frac{\Vert \mathbf{\Phi}_{l}^{-\frac{1}{2}}(\hat{\mathbf{m}}_{l}-\rho_{a,l}\hat{\mathbf{b}}_{l}-\nu_{l}\mathbf{1}) \Vert^2}{2(1-\beta_{1,l})} \leq  \frac{3\Vert \mathbf{\Phi}_{l}^{-\frac{1}{2}}\hat{\mathbf{m}}_{l}\Vert^2}{2(1-\beta_{1})}
		+\frac{3\rho_{a,l}^2\Vert \mathbf{\Phi}_{l}^{-\frac{1}{2}}\hat{\mathbf{b}}_{l}
			\Vert^2}{2(1-\beta_{1})} + \frac{3\nu_{l}^2\Vert \mathbf{\Phi}_{l}^{-\frac{1}{2}}\mathbf{1}
			\Vert^2}{2(1-\beta_{1})}$. 
		Then, the upperbound of (\ref{grieq00}) can be  given by
		\begin{align}
			&\left<\mathbf{d}_{u}^{(k,l)},\Delta \mathbf{u}^{(k,l)}\right>{\leq}  \underbrace{\frac{\Vert \mathbf{\Phi}_{l}^{\frac{1}{2}}\Delta \mathbf{u}^{(k,l)}\Vert^2}{2(1-\beta_{1,l})} -\frac{\Vert \mathbf{\Phi}_{l}^{\frac{1}{2}}\Delta \mathbf{u}^{(k,l+1)}\Vert^2}{2(1-\beta_{1,l})}}_{\text{\ding{172}}}\notag\\
			&+\underbrace{\frac{\beta_{1,l}\Vert \mathbf{\Phi}_{l}^{-\frac{1}{2}}\hat{\mathbf{m}}_{l-1}
					\Vert^2}{2(1-\beta_{1})}}_{\text{\ding{173}}}+ \underbrace{\frac{\rho_{a,l}\Vert \mathbf{\Phi}_{l}^{-\frac{1}{2}}\hat{\mathbf{b}}_{l}
					\Vert^2}{2(1-\beta_{1})}}_{\text{\ding{174}}}\notag + \underbrace{ \frac{\nu_{l}\Vert \mathbf{\Phi}_{l}^{-\frac{1}{2}}\mathbf{1}
					\Vert^2}{2(1-\beta_{1})} }_{\text{\ding{175}}}\\
			&+ \underbrace{\frac{\beta_{1,l}\Vert \mathbf{\Phi}_{l}^{\frac{1}{2}}\Delta \mathbf{u}^{(k,l)}\Vert^2}{2(1-\beta_{1})}}_{\text{\ding{176}}}+ \underbrace{ \frac{\rho_{a,l}\Vert \mathbf{\Phi}_{l}^{\frac{1}{2}}\Delta \mathbf{u}^{(k,l)}\Vert^2}{2(1-\beta_{1})} }_{\text{\ding{177}}}
			+\underbrace{\frac{\nu_{l}\Vert \mathbf{\Phi}_{l}^{\frac{1}{2}}\Delta \mathbf{u}^{(k,l)}\Vert^2}{2(1-\beta_{1})} }_{\text{\ding{178}}}\label{grieq}\notag\\
			&+\underbrace{ \frac{3\Vert \mathbf{\Phi}_{l}^{-\frac{1}{2}}\hat{\mathbf{m}}_{l}			 \Vert^2}{2(1-\beta_{1})} }_{\text{\ding{179}}}
			+\underbrace{ \frac{3\rho_{a,l}^2\Vert \mathbf{\Phi}_{l}^{-\frac{1}{2}}\hat{\mathbf{b}}_{l}
					\Vert^2}{2(1-\beta_{1})} }_{\text{\ding{180}}}+\underbrace{ \frac{3\nu_{l}^2\Vert \mathbf{\Phi}_{l}^{-\frac{1}{2}}\mathbf{1}
					\Vert^2}{2(1-\beta_{1})} }_{\text{\ding{181}}},\notag
		\end{align}
		
		To bound $R_L$, we upperbound of the summation of the terms \ding{172}-\ding{181} over the index $l$ as follows.

		\subsubsection{Term \ding{172}}
		It can be shown that
		\begin{equation}\label{beta1lPhil}
			\begin{split}
				&\Vert \mathbf{\Phi}_{l}^{\frac{1}{2}}\Delta \mathbf{u}^{(k,l)}\Vert^2=\sum_{i=1 }^{N}\phi_{l,i}^{-1} |\Delta u_{i}^{(k,l)}|^2\\
				&=\frac{1}{\alpha_{1,l}}\sum_{i=1 }^{N}\sqrt{|\hat{v}_{l,i}|}\cdot |\Delta u_{i}^{(k,l)}|^2 +\rho_{a,l}
				||\Delta \mathbf{u}^{(k,l)}||^2\\
				&\overset{(a)}{=}\frac{1}{\alpha_{1,l}}\sum_{i=1}^{N} \sum_{p=1}^{l} \sqrt{1-\beta_{2}}\beta_{2}^{\frac{l-p}{2}} |d_{u,i}^{(k,p)}| \cdot|\Delta u_{i}^{(k,l)}|^2\\
				&\quad+\rho_{a,l}
				||\Delta \mathbf{u}^{(k,l)}||^2\\
				&\leq  \frac{\sqrt{1-\beta_{2}}}{\alpha_{1}^{-}(1-\sqrt{\beta_2})} D_{u,1} D_{\Delta} \sqrt{l} + D_{\Delta}  \rho_{a} \eta_{a}^l,
			\end{split}
		\end{equation}
		where step (a) comes from (\ref{hatvl}).
		Then, with the decreasing learning rate $\phi_{l,i}^{-1}$, we have
		\begin{equation}\label{summationterm}
			\begin{split}
				&\sum_{l=1}^{L}\left(\frac{\Vert \mathbf{\Phi}_{l}^{\frac{1}{2}}\Delta \mathbf{u}^{(k,l)}\Vert^2}{2(1-\beta_{1,l})} -\frac{\Vert \mathbf{\Phi}_{l}^{\frac{1}{2}}\Delta \mathbf{u}^{(k,l+1)}\Vert^2}{2(1-\beta_{1,l})}\right)\\
				&\leq\frac{\Vert \mathbf{\Phi}_{1}^{\frac{1}{2}}\Delta \mathbf{u}^{(k,1)}\Vert^2}{2(1-\beta_{1})}+\frac{\Vert \mathbf{\Phi}_{L}^{\frac{1}{2}}\Delta \mathbf{u}^{(k,L+1)}\Vert^2}{2(1-\beta_{1})}\\
				&\quad +\sum_{l=2}^{L}\left(\frac{\Vert \mathbf{\Phi}_{l}^{\frac{1}{2}}\Delta \mathbf{u}^{(k,l)}\Vert^2}{2(1-\beta_{1,l})} -\frac{\Vert \mathbf{\Phi}_{l-1}^{\frac{1}{2}}\Delta \mathbf{u}^{(k,l)}\Vert^2}{2(1-\beta_{1,l})}\right)\\
				&\overset{(a)}{\leq} \frac{\sqrt{1-\beta_{2}}D_{u,1} D_{\Delta}}{\alpha_{1}^{-}(1-\sqrt{\beta_2})(1-\beta_{1})}  \sqrt{L} +  \frac{\rho_{a} \eta_{a}^{L} D_{\Delta}}{1-\beta_{1}}\\
				&\quad + \frac{\sqrt{1-\beta_{2}}D_{u,1} D_{\Delta}}{\alpha_{1}^{-}(1-\sqrt{\beta_2})(1-\beta_{1})}   +  \frac{\rho_{a} \eta_{a} D_{\Delta}}{1-\beta_{1}}\\
				&\quad +  \frac{\sum\limits_{l=2}^{L}\sum\limits_{i=1}^{N}  \left(\phi_{l,i}-\phi_{l-1,i} \right)|\Delta u_{i}^{(k,l)}|^2}{2(1-\beta_{1})} \\
				&\overset{(b)}{\leq} \frac{\sqrt{1-\beta_{2}}D_{u,1} D_{\Delta}}{\alpha_{1}^{-}(1-\sqrt{\beta_2})(1-\beta_{1})}  \sqrt{L} +  \frac{\rho_{a} \eta_{a} D_{\Delta}}{1-\beta_{1}}\\
				&\quad + \frac{\sqrt{1-\beta_{2}}D_{u,1} D_{\Delta}}{\alpha_{1}^{-}(1-\sqrt{\beta_2})(1-\beta_{1})}   +  \frac{\rho_{a} \eta_{a} D_{\Delta}}{1-\beta_{1}} + \frac{D_{\Delta_{u,2}}D_{\phi}}{1-\beta_1},
			\end{split}
		\end{equation}
		where step (a) follows  (\ref{beta1lPhil}) and step (b) follows 
		\begin{equation}
			\begin{split}
				&\sum\limits_{l=2}^{L}\sum\limits_{i=1}^{N}  \left(\phi_{l,i}-\phi_{l-1,i} \right)|\Delta u_{i}^{(k,l)}|^2\\
				&\leq D_{\Delta u, 2}\sum\limits_{i=1}^{N} \sum\limits_{l=2}^{L}  \left(\phi_{l,i}-\phi_{l-1,i} \right)\leq 2D_{\Delta_{u,2}}D_{\phi}.
			\end{split}
		\end{equation}
		

		\subsubsection{Terms \ding{173} $\&$ \ding{179}}
		Since $(1-\beta_{1})$ is a non-zero constant, we focus on the upperbound of the terms $\sum_{l=1}^{L}\Vert \mathbf{\Phi}_{l}^{-\frac{1}{2}}\hat{\mathbf{m}}_{l}\Vert^2$ and $\sum_{l=1}^{L}\beta_{1,l}\Vert \mathbf{\Phi}_{l}^{-\frac{1}{2}}\hat{\mathbf{m}}_{l-1}\Vert^2$. 
		Denote $\hat{m}_{l,i}$ and $d_{u,i}$ as the $i$th entry of $\hat{\mathbf{m}}_{l}$ and $\mathbf{d}_u^{(k,l)}$, respectively. Then, we have
		\begin{equation}\label{ineq1}
			\begin{split}
				&\Vert \mathbf{\Phi}_{l}^{-\frac{1}{2}}\hat{\mathbf{m}}_{l}\Vert^2=\sum_{i=1 }^{N}\frac{\hat{m}_{l,i}^2}{\phi_{l,i}}\leq \sum_{i=1 }^{N}\frac{\hat{m}_{l,i}^2}{\sqrt{|\hat{v}_{l,i}|}/\alpha_{1,l}}\\
				&= \sum_{i=1 }^{N}\frac{ \left(\sum_{p=1}^{l}(1-\beta_{1,p})\prod_{q=1}^{l-p}\beta_{1,l-q+1}  d_{u,i}^{(k,p)} \right)^2   }{\sqrt{|\hat{v}_{l,i}|}/\alpha_{1,l}}\\
				&\overset{(a)}{\leq}\sum_{i=1 }^{N}\frac{\alpha_{1,l}\eta_{1}^{2l}\left(\sum_{p=1}^{l}\beta_{1}^{l-p} \right)  \left(\sum_{p=1}^{l}\beta_{1}^{l-p} (d_{u,i}^{(k,p)})^2  \right)  }{\sqrt{\sum_{p=1}^{l}(1-\beta_{2})\beta_{2}^{l-p} |d_{u,i}^{(k,p)}|^2}}\\
				&\overset{(b)}{\leq} \frac{\alpha_{1,l}\eta_{1}^{2l}}{(1-\beta_{1})\sqrt{1-\beta_{2}}}\sum_{i=1 }^{N}  \left(\sum_{p=1}^{l}
				\left(\frac{\beta_{1}}{\sqrt{\beta_{2}}}\right)^{l-p}
				|d_{u,i}^{(k,p)}|  \right),\\
			\end{split}
		\end{equation}
		where step (a) comes from the inequality $(1-\beta_{1,p})\leq 1$,  $\prod_{q=1}^{l-p}\beta_{1,l-q+1}\leq \beta_{1}^{l-p}\eta_{1}^l$ and the Jensen inequality, i.e., $\left(\frac{\sum_i a_i b_i}{\sum_i a_i}\right)^2\leq \frac{\sum a_i b_i^2}{\sum_i a_i}$, and step (b) follows the inequalities  $\sum_{p=1}^{l}\beta_{1}^{l-p}\leq \frac{1}{1-\beta_1}$ and $\sum_{p=1}^{l}(1-\beta_{2})\beta_{2}^{l-p} |d_{u,i}^{(k,p)}|^2 \geq (1-\beta_{2})\beta_{2}^{l-p} |d_{u,i}^{(k,p)}|^2$.

		By summing up \eqref{ineq1} over the index $l$, we have
		\begin{align}
			&\sum_{l=1}^L\Vert \mathbf{\Phi}_{l}^{-\frac{1}{2}}\hat{\mathbf{m}}_{l}\Vert^2\notag\\
			&\leq\sum_{l=1}^L\frac{\alpha_{1,l}\eta_{1}^{2l}}{(1-\beta_{1})\sqrt{1-\beta_{2}}}\sum_{i=1 }^{N}  \left(
			\sum_{p=1}^{l}
			\left(\frac{\beta_{1}}{\sqrt{\beta_{2}}}\right)^{l-p}
			|d_{u,i}^{(k,p)}|  \right)\notag\\
			&=\sum_{l=1}^L\frac{\alpha_{1,l}\eta_{1}^{2l}}{(1-\beta_{1})\sqrt{1-\beta_{2}}} ||\mathbf{d}_{u}^{(k,l)}||_1  \left(\sum_{j=l}^{L}
			\left(\frac{\beta_{1}}{\sqrt{\beta_{2}}}\right)^{j-l}
			\right)\\
			&\leq \frac{\alpha_{1}^{+}D_{u,1}}{(1-\beta_{1})(1-\frac{\beta_{1}}{\sqrt{\beta_{2}}})\sqrt{1-\beta_{2}}}  \sum_{l=1}^L\frac{\eta_{1}^{2l}}{\sqrt{l}} \notag\\
			&\overset{(a)}{\leq}    \frac{\alpha_{1}^{+}D_{u,1}}{(1-\beta_{1})(1-\frac{\beta_{1}}{\sqrt{\beta_{2}}})\sqrt{1-\beta_{2}}(1-\eta_{1}^2)},\notag
		\end{align}
		where 
		we have utilized the property that $\sum_{l=1}^L\frac{\eta_{1}^{2l}}{\sqrt{l}} \leq \sum_{l=1}^L \eta_{1}^{2l} \leq \frac{1}{1-\eta_{1}^2}$ in step (a).  
		Then, we have
		\begin{equation}\label{phil12ml}
			\begin{split}
				&\sum_{l=1}^{L}\Vert \mathbf{\Phi}_{l}^{-\frac{1}{2}}\hat{\mathbf{m}}_{l}\Vert^2\leq \frac{\alpha_{1}^{+}D_{u,1}}{(1-\beta_{1})(1-\frac{\beta_{1}}{\sqrt{\beta_{2}}})\sqrt{1-\beta_{2}}(1-\eta_{1}^2)}. \notag
			\end{split}
		\end{equation}
		
		Similarly, we can obtain 
		\begin{equation}\label{corofirsttermeq}
			\begin{split}
				&\sum_{l=1}^{L}\beta_{1,l}\Vert \mathbf{\Phi}_{l}^{-\frac{1}{2}}\hat{\mathbf{m}}_{l-1}\Vert^2 \leq \sum_{l=1}^{L}\beta_{1,l}\Vert \mathbf{\Phi}_{l-1}^{-\frac{1}{2}}\hat{\mathbf{m}}_{l-1}\Vert^2\\
				&\leq \frac{\alpha_{1}^{+}\beta_{1}D_{u,1}}{(1-\beta_{1})(1-\frac{\beta_{1}}{\sqrt{\beta_{2}}})\sqrt{1-\beta_{2}}(1-\eta_{1}^2)}.
			\end{split}
		\end{equation}

		\subsubsection{Terms \ding{174} $\&$ \ding{180}}
		First, we have 
		\begin{equation}
			\begin{split}
				&\sum_{l=1}^{L}\rho_{a,l}\Vert \mathbf{\Phi}_{l}^{-\frac{1}{2}}\hat{\mathbf{b}}_{l}	\Vert^2 \leq \sum_{l=1}^{L}\rho_a \eta_a^{l} D_{\phi}^{l} ||\hat{\mathbf{b}}_{l}||^2 	\leq \frac{\rho_a  D_{\phi} D_{b,2}}{1-\eta_a},\notag
			\end{split}
		\end{equation}
		where $D_{\phi}^{l}= \max\limits_i \phi_{l,i}^{-1}$. 
		Similarly, we can obtain
		\begin{equation}\label{corosecondtermeq}
			\sum_{l=1}^{L}\rho_{a,l}^2\Vert \mathbf{\Phi}_{l}^{-\frac{1}{2}}\hat{\mathbf{b}}_{l}
			\Vert^2	\leq \frac{\rho_a^2  D_{\phi} D_{b,2}}{1-\eta_a^2}.
		\end{equation}

		\subsubsection{Terms \ding{175} $\&$ \ding{181}} By the definition of $\nu_{l}$ in (\ref{nuldef}), we have
		\begin{equation}
			\begin{split}
				\nu_{l}&
				\leq||\mathbf{d}_{\star}^{(k,l)}||_1\leq ||\hat{\mathbf{m}}_{l}||_1 + \rho_{a,l} ||\hat{\mathbf{b}}_{l}||_1.
			\end{split}		
		\end{equation}
		Similar to (\ref{ineq1}), we can obtain 
		\begin{equation}\label{lemmanusum1}
			\begin{split}
				&\sum_{l=1}^{L}||\hat{\mathbf{m}}_{l}||_1
				\leq \sum_{l=1}^{L}\sum_{i=1 }^{N} \left(\sum_{p=1}^{l} \prod_{q=1}^{l-p}\beta_{1,l-q+1}  |d_{u,i}^{(k,p)}|  \right)  \\
				&
				\leq \sum_{l=1}^{L}\frac{ ||\mathbf{d}_{u}^{(k,p)}||_1 \eta_1^l}{(1-\beta_1)}\leq \frac{ D_{u,1}}{(1-\eta_1)(1-\beta_1)}.
			\end{split}		
		\end{equation}
		Then, we have
		\begin{equation}\label{lemmanusum2}
			\begin{split}
				&	\sum_{l=1}^L\rho_{a,l} ||\hat{\mathbf{b}}_{l}||_1=\rho_{a}D_{b,1}\sum_{l=1}^L  \eta_{a}^{l}\leq \frac{\rho_{a}D_{b,1}}{1-\eta_{a}}.\\
			\end{split}		
		\end{equation}
		
		By substituting (\ref{lemmanusum1}) and (\ref{lemmanusum2}) into (\ref{nulemma}), we have
		\begin{equation}
			\begin{split}
				&\sum_{l=1}^{L}\nu_{l}\Vert \mathbf{\Phi}_{l}^{-\frac{1}{2}}\mathbf{1} \Vert^2 \leq D_{\phi}\sum_{l=1}^{L}\nu_{l}\\
				&\quad \quad \leq D_{\phi}\left( \frac{ D_{u,1}}{(1-\eta_1)(1-\beta_1)}+\frac{\rho_{a}D_{b,1}}{1-\eta_{a}}\right).
			\end{split}
		\end{equation}
		
		It thus follows that 
		\begin{equation}\label{nulemma}
			\begin{split}
				&\sum_{l=1}^{L}\nu_{l}\Vert \mathbf{\Phi}_{l}^{-\frac{1}{2}}\mathbf{1} \Vert^2 \leq  \frac{ D_{u,1}D_{\phi}}{(1-\eta_1)(1-\beta_1)}+\frac{\rho_{a}D_{b,1}D_{\phi}}{1-\eta_{a}}.
			\end{split}
		\end{equation}
		
		Similarly, we can obtain
		\begin{equation}\label{nulemma2}
			\begin{split}
				&\sum_{l=1}^{L}\nu_{l}^2\Vert \mathbf{\Phi}_{l}^{-\frac{1}{2}}\mathbf{1} \Vert^2 \leq  \frac{2 D_{u,1}^2 D_{\phi}}{(1-\eta_1^2)(1-\beta_1)^2}+\frac{2\rho_{a}^2D_{b,1}^2D_{\phi}}{(1-\eta_{a}^2)}.\notag
			\end{split}
		\end{equation}
		
		\subsubsection{Term \ding{176}} 
		By \eqref{beta1lPhil}, we have 
		\begin{equation}
			\begin{split}\label{lemmabeta1l}
				&\sum_{l=1}^{L}\beta_{1,l}\Vert \mathbf{\Phi}_{l}^{\frac{1}{2}}\Delta \mathbf{u}^{(k,l)}\Vert^2\\
				&\leq  \frac{\beta_{1}\sqrt{1-\beta_{2}}}{\alpha_{1}^{-}(1-\sqrt{\beta_2})} D_{u,1} D_{\Delta} \sqrt{l}\eta_{1}^l + D_{\Delta} \beta_{1} \rho_{a} \eta_{1}^l\eta_{a}^l\\
				&\overset{(a)}{\leq} \frac{\beta_{1}\sqrt{1-\beta_{2}}D_{u,1} D_{\Delta}}{\alpha_{1}^{-}(1-\sqrt{\beta_2})(1-\eta_{1})^2}  + \frac{\beta_{1} \rho_{a}D_{\Delta}}{(1-\eta_{1}\eta_{a})},
			\end{split}
		\end{equation}
		where we have utilized the bound of the arithmetic-geometric series, i.e., $\sum_{l=1}^{L} l \eta_1^l\leq \frac{1}{(1-\eta_1)^2}$ in (a).

		\subsubsection{Term \ding{177}}
		
		By replacing $\beta_{1,l}$ with $\rho_{a,l}$  in (\ref{lemmabeta1l}), we have
		\begin{equation}\label{cororaoal}
			\begin{split}
				&\sum_{l=1}^{L}\rho_{a,l}\Vert \mathbf{\Phi}_{l}^{\frac{1}{2}}\Delta \mathbf{u}^{(k,l)}\Vert^2\leq \frac{\rho_{a}\sqrt{1-\beta_{2}}D_{u,1} D_{\Delta}}{\alpha_{1}^{-}(1-\sqrt{\beta_2})(1-\eta_{a})^2}  + \frac{\rho_{a}^2 D_{\Delta}}{1-\eta_{a}^2}.\notag
			\end{split}
		\end{equation}
		
		\subsubsection{Term \ding{178}}
		
		Recalling (\ref{lemmanusum1}) and (\ref{lemmanusum2}), we have 
		\begin{equation}\label{lemmanul}
			\begin{split}	
				\nu_{l}&\leq ||\hat{\mathbf{m}}_{l}||_1 + \rho_{a,l} ||\hat{\mathbf{b}}_{l}||_1\leq \frac{ D_{u,1}}{(1-\beta_1)}\eta_1^l+ \rho_{a}D_{b,1} \eta_{a}^{l}.
			\end{split}
		\end{equation}
		Then, we can obtain  
		\begin{equation}\label{nulvert}
			\begin{split}
				&\nu_{l}\Vert \mathbf{\Phi}_{l}^{\frac{1}{2}}\Delta \mathbf{u}^{(k,l)}\Vert^2=\nu_{l}\sum_{i=1 }^{N}\phi_{l,i} |\Delta u_{i}^{(k,l)}|^2\\
				&=\frac{\nu_{l}}{\alpha_{1,l}}\sum_{i=1 }^{N}\sqrt{|\hat{v}_{l,i}|}\cdot |\Delta u_{i}^{(k,l)}|^2 +\nu_{l}\rho_{a,l}
				||\Delta \mathbf{u}^{(k,l)}||^2\\
				&\leq  \frac{ \nu_{l}\sqrt{1-\beta_{2}}}{\alpha_{1}^{-}(1-\sqrt{\beta_2})} D_{u,1} D_{\Delta} \sqrt{l}+ \nu_{l} D_{\Delta}  \rho_{a} \eta_{a}^l .
			\end{split}
		\end{equation}
		
		Similarly, we have
		\begin{equation}\label{sumsqrtlnu}
			\begin{split}
				\sum_{l=1}^L \sqrt{l} \nu_{l} &=
				\sum_{l=1}^L \sqrt{l} \left(\frac{ D_{u,1}}{(1-\beta_1)}\eta_1^l+ \rho_{a}D_{b,1} \eta_{a}^{l} \right) \\
				&\leq \frac{ D_{u,1}}{(1-\beta_1)(1-\eta_1)^2}+  \frac{\rho_{a}D_{b,1}}{(1-\eta_a)^2},
			\end{split}
		\end{equation}
		\begin{equation}\label{sumetaanu}
			\begin{split}
				\sum_{l=1}^L \eta_{a}^l \nu_{l} &=
				\sum_{l=1}^L \eta_{a}^l \left(\frac{ D_{u,1}}{(1-\beta_1)}\eta_1^l+ \rho_{a}D_{b,1} \eta_{a}^{l} \right) \\
				&\leq \frac{ D_{u,1}}{(1-\beta_1)(1-\eta_1)(1-\eta_{a})}+  \frac{\rho_{a}D_{b,1}}{(1-\eta_a)^2}.
			\end{split}
		\end{equation}
		By substituting (\ref{sumsqrtlnu}) and (\ref{sumetaanu}) into (\ref{nulvert}), we have
		\begin{align}
			&\sum_{l=1}^L \nu_{l}\Vert \mathbf{\Phi}_{l}^{\frac{1}{2}}\Delta \mathbf{u}^{(k,l)}\Vert^2\notag\\
			&\leq \left(\frac{ D_{u,1}}{(1-\beta_1)(1-\eta_1)^2}+  \frac{\rho_{a}D_{b,1}}{(1-\eta_a)^2}\right)\frac{\sqrt{1-\beta_{2}}D_{u,1} D_{\Delta}}{\alpha_{1}(1-\sqrt{\beta_2})} \notag\\
			&+  \left(\frac{ D_{u,1}}{(1-\beta_1)(1-\eta_1)(1-\eta_{a})}+  \frac{\rho_{a}D_{b,1}}{(1-\eta_a)^2}\right)D_{\Delta}  \rho_{a}.\label{lemmdanuphideltaul}
		\end{align}

		By combining the upperbounds for the summations of terms $\text{\ding{172}}$-$\text{\ding{181}}$, (\ref{regretbound}) can be proved.
		


	\end{document}